\documentclass[
 reprint,
 amsmath,amssymb,
 aps,
]{revtex4-2}

\usepackage{graphicx}
\usepackage{dcolumn}
\usepackage{bm}
\usepackage{natbib}

\begin{document}

\preprint{APS/123-QED}

\title{Energy-information trade-off makes the cortical critical power law the optimal coding}

\author{Tsuyoshi Tatsukawa}
\author{Jun-nosuke Teramae}
\affiliation{%
 Graduate School of Informatics, Kyoto University, Sakyo-ku, Kyoto 606-8502, Japan\\
}

\begin{abstract}

\end{abstract}

\begin{abstract}
Stimulus responses of cortical neurons exhibit the critical power law, where the covariance eigenspectrum follows the power law with the exponent just at the edge of differentiability of the neural manifold. This criticality is conjectured to balance the expressivity and robustness of neural codes, because a non-differential fractal manifold spoils coding reliability. However, contrary to the conjecture, here we prove that the neural coding is not degraded even on the non-differentiable fractal manifold, where the coding is extremely sensitive to perturbations. Rather, we show that the trade-off between energetic cost and information always makes this critical power-law response the optimal neural coding. Direct construction of a maximum likelihood decoder of the power-law coding validates the theoretical prediction. By revealing the non-trivial nature of high-dimensional coding, the theory developed here will contribute to a deeper understanding of criticality and power laws in computation in biological and artificial neural networks.
\end{abstract}

\maketitle

\section{Introduction}\label{sec1}

How the activity of a population of neurons in the brain represents, or encodes, external signals such as visual images presented to animals has long been a central question in both neuroscience and machine learning \cite{barlow1961possible,Simoncelli2001-ar,Churchland2012-iu,Rigotti2013-oy,Sadtler2014-uf,Chung2018-og,Gallego2018-qq,Stringer2019-dw,Gallego2020-ap,Yoshida2020-ji,Chung2021-ao,Sorscher2022-da,Rust2022-av,Nogueira2023-se,Schneider2023-cj,Manley2024-bn}. While the actual structure of the coding has been largely unknown, two seemingly contradictory hypotheses have been proposed. One is the efficient coding hypothesis which implies that the population coding should be high-dimensional and sparse in order to reduce the correlation of input stimuli to make decoding easier\cite{barlow1961possible,Simoncelli2001-ar,Rigotti2013-oy,Manley2024-bn}. The other is the low-dimensional subspace hypothesis, which suggests that the population activities should be confined to low-dimensional subspaces or manifolds to make the coding redundant and robust to noise\cite{Churchland2012-iu,Sadtler2014-uf,Gallego2018-qq,Gallego2020-ap,Nogueira2023-se,Schneider2023-cj}.

\begin{figure*}[!tb]
 \centering
 \includegraphics[width=16cm]{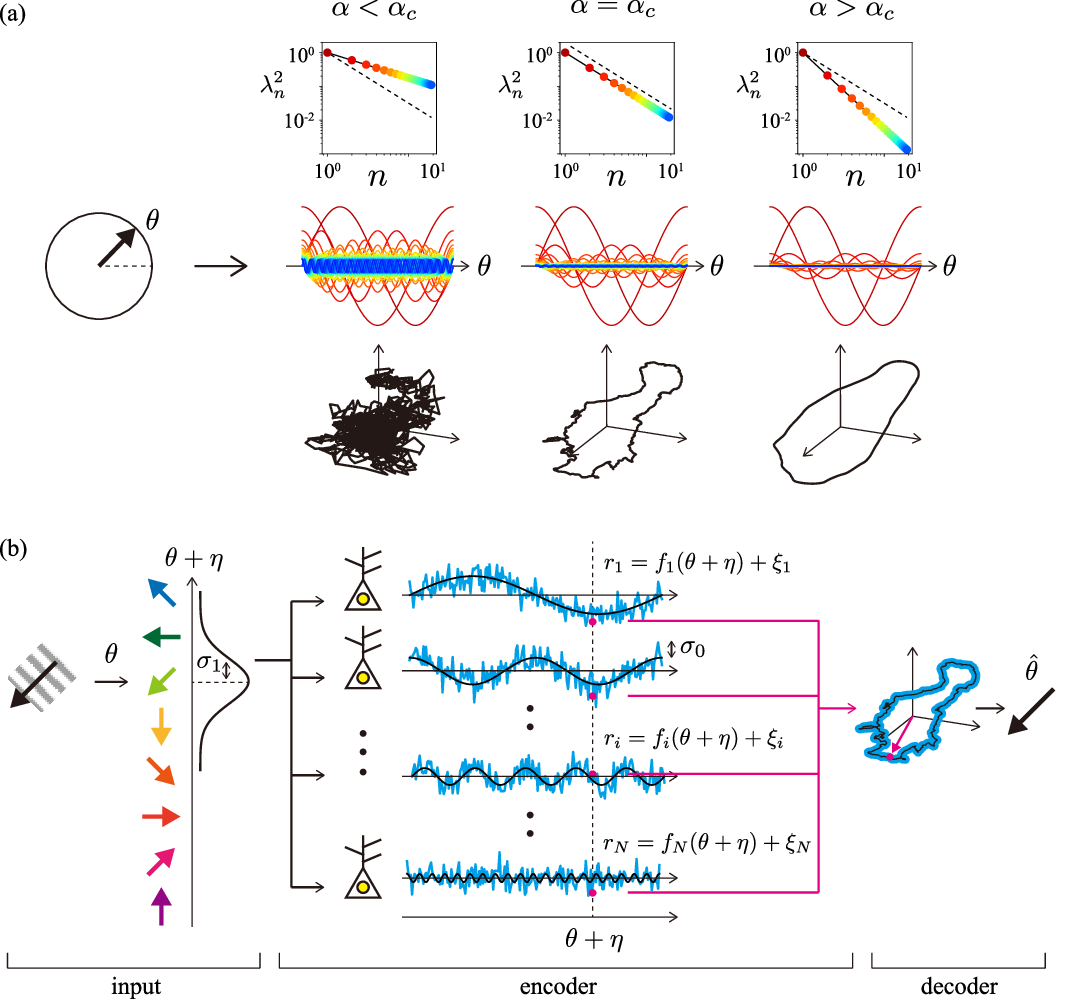}
 \caption{Power-law responses of cortical neurons and an analytically tractable coding model with the power-law responses. (a) The variance of the $n$th principal component, i.e., the square of the eigenspectrum of the covariance matrix, of population activities of cortical neurons in response to input stimuli follows the power law with $\lambda_n^2 \propto k^{-\alpha}$ (upper rows). If the exponent $\alpha$ is smaller than the critical value $\alpha_c$, the neural manifold of the stimulus representation will be high-dimensional and non-differentiable (left column). In contrast, the manifold will be low-dimensional and smooth if the value of $\alpha$ is larger than $\alpha_c$ (right column). In this sense, the experimentally observed exponent close to $\alpha_c$ is critical (middle column). Panels of the middle row show examples of receptive fields of neurons (different colors corresponding to different neurons) where the input stimulus $\theta$ is a one-dimensional periodic signal, and neural responses obey each power law. Lower panels depict 3-dimensional projections of the neural manifolds, that is, projections of neural responses shown in the middle rows to randomly chosen three directions when the input $\theta$ varies from $0$ to $2\pi$. (b) Outline of the encoding model with the power-law stimulus responses. Input stimulus $\theta$ (furthest to the left) added an intrinsic noise $\eta$ with strength $\sigma_1$ is given to the population of neurons. Each neuron responds to the input with the activity following its receptive field whose amplitude obeys the power law $f_i(\theta+\eta)$, which is further added random noise $\xi$ of the strength $\sigma_0$ which represents the randomness of the neural activity. Thus, the population activity is given as a perturbed point on the neural manifold. Finally, the output neuron, or the estimator (furthest to the right), decodes the population activity to obtain the estimated value $\hat{\theta}$ of the input.}
 \label{fig:fig1}
\end{figure*}

Recently, the actual structure of the population coding in the brain has been revealed owing to the rapid development of recording techniques of neural activities \cite{Stringer2019-dw}. This shows that actual coding in the cortex is in the middle point between the two hypotheses. Simultaneous recording of a large number of neurons in the primary visual cortex in vivo shows that the eigenspectrum of the covariance matrix of the neural activity marginalized over the input stimuli follows the power law as the variance of the $n$th dimension of the population activity decays with the power of $n$, in almost ascending order of its wavenumber or frequency of the response to the input (Fig. \ref{fig:fig1}a). The exponent of the power-law decay agrees well with $\alpha_c = 1 + 2/D$, regardless of the input statistics, where $D$ is the dimension of the input. Therefore, the eigenspectrum universally decays as $1/n$ for natural images because the dimension of the input images is almost infinitely large. The efficient coding hypothesis predicts that the spectrum must be almost flat rather than decaying, and the low-dimensional subspace hypothesis does that it should decay rapidly, the observed power-law decay therefore implies that the brain is indeed in the midpoint between them.

The authors of the above paper also proved that the exponent $\alpha_c$ is the critical value as the exponent gives the border of differentiability of the coding manifold, and therefore the brain is in the so-called ``edge of chaos''. They proved mathematically that, in the limit of large number of neurons, if the map defined by the coding from the input space to the space of the neural activity is differentiable, i.e., its neural manifold is differentiable, the eigenspectrum should asymptotically decay faster than the critical exponent $\alpha_c$. Conversely, if the eigenspectrum decays slower than the exponent, the map must be non-differentiable, and the neural response will be fractal, which means that the coding must be too sensitive to input perturbations because even infinitesimally close two inputs elicit significantly different neural activities. It is then conjectured that the brain strikes a balance between the expressiveness and robustness of the coding, because the critical power-law coding is as sensitive to inputs as possible while maintaining its smoothness. The similar critical sensitivity to perturbations also plays an important role in studies of machine learning \cite{Poole2016-wo,Schoenholz2016-ll,Pennington2017-hk,Kaplan2020-zt,Nassar2020-pz,Chen2022-ts,Johnston2023-ov,Joseph2023-ct}.

It is truly fascinating that the brain's signal encoding is in a critical state on the edge of chaos\cite{Beggs2003-ac,De_Arcangelis2006-wx,Tkacik2015-ly,Dahmen2019-qr,Fosque2021-fw,OByrne2022-no,Morales2023-ii}. However, quantitative evaluation of the ability of the power-law coding remain elusive probably due to the lack of the framework to describe coding ability of the power-law theoretically. For example, it remains unclear how the non-differentiability of the neural manifold indeed affect the coding performance, and why the critical state is used by the brain as an encoder.

Here, we address this problem by adopting the framework of statistical parameter estimation to the power-law coding with properly accounting for possible noise sources of it \cite{Seung1993-qt,Brunel1998-wb,Deneve1999-sf,Sompolinsky2001-dl,Quian_Quiroga2009-rz,Moreno-Bote2014-dh}. In this framework, a desirable neural coding is defined as the one that allow decoders to estimate its encoded signals with minimal estimation error. This ability of the stochastic coding is measured by its Fisher information because the inverse of the Fisher information of the stochastic coding gives the lower bound of the variance of the estimation error for unbiased decoders (Cramer-Rao bound) and the asymptotic variance of the error for maximum likelihood decoders in the large number limit of observations \cite{Brunel1998-wb,cover1999elements}.

We will derive analytical expressions for the Fisher information of the neural power law coding and prove that (i) the exponent $\alpha_c$ does indeed give the critical point in the sense that its Fisher information is discontinuous only at this exponent in the limit of the large numbers of neurons. Unexpectedly, however, we will see that (ii) the Fisher information is kept constant rather than decreases even for the exponent smaller than the critical value and thus for the case where its coding manifold is non-differentiable. Thus, contrary to the previous conjecture, the critical power law is not always the best coding in the information-theoretic sense. So why does the brain use the critical coding? We will show that (iii) the introduction of the energetic cost ensures that the critical coding always achieves optimality. In the derivation of Fisher information, we will also prove the remarkable relationship between the variance and the susceptibility of neural activity, corresponding to the fluctuation-dissipation theorem in statistical physics \cite{Kubo1966-rg}. To confirm the theoretical predictions and to verify the optimality of the critical coding, we directly construct a maximum likelihood decoder for the power-law coding by explicitly integrating the conditional probability distribution of the encoded input signal. This allows us to measure the variance of its estimation errors and to numerically evaluate the Fisher information of the power-law coding.

\section{Results}\label{sec2}

\subsection{Statistical estimation with the power-law population coding}

Following Stringer's pioneering work, let us start from the simplest case where a population of neurons encodes a one-dimensional periodic scalar angular variable, $\theta \in [0, 2\pi)$, such as the orientation of a line segment presented in animal's visual field, thus $D=1$ (Fig. \ref{fig:fig1}b). Generalizations to higher input dimensions will be given later. This coding can be affected by two different noise sources; one is the input noise that is put directly to the input stimulus $\theta$ and the other is the neural noise being independently put on each neural activity evoked by the input stimulus. We model the input noise $\eta$ and the neural noise for the $i$th neuron $\xi_i$ using the independent Gaussian random variables satisfying that $\langle \xi_i \rangle = \langle \eta \rangle = \langle \eta \xi_i \rangle = 0$, $\langle \eta^2 \rangle = \sigma_1^2$, and $\langle \xi_i \xi_j \rangle = \sigma_0^2 \delta_{ij} $, where $\sigma_1$ and $\sigma_0$ are the strengths of the input and neural noise, respectively. Then, based on the experimental results that the variance of the principal components, i.e., the eigenspectrum of the covariance matrix of the neural activities, decays with the power law in ascending order of its frequency, we express the neural activity of the $i$th neuron ($i=1, \ldots, 2N$) as an expansion over the orthonormal Fourier basis of a scalar periodic function,
\begin{align*}
    r_i(\theta) = \sum_{j=1}^{2N} a_{ij} f_j(\theta + \eta) + \xi_i,
\end{align*}
where
\begin{align*}
    f_j(\theta) = \begin{cases}
        c n^{-\alpha/2} \cos{n \theta} & (j=2n-1)\\
        c n^{-\alpha/2} \sin{n \theta} & (j=2n)
    \end{cases}.
\end{align*}
Here, $A = [a_{ij}]$ is an orthogonal matrix that rotates the axes of the basis function to the axes of the neuron space, and $c$ is a scaling factor that determines the magnitude of the activity of the neurons. However, by using a basis change, we can set $A$ to be the identity matrix and $c=1$ without loss of generality, which gives the stochastic coding model as,
\begin{align}
    r_i(\theta) = \begin{cases}
        n^{-\alpha/2} \cos{n(\theta + \eta)} + \xi_i & (i=2n-1)\\
        n^{-\alpha/2} \sin{n(\theta + \eta)} + \xi_i & (i=2n)
    \end{cases}. \label{eq:coding}
\end{align}
Note that the axis change does not affect the strength of the neural noise since the noise is isotropic in the neural space.

The Fisher information of the stochastic coding is defined as the variance of the score function, the derivative of the loglikelihood function with respect to the input $\theta$, or equivalently, the negative mean of the second derivative of the loglikelihood function \cite{Brunel1998-wb,cover1999elements}
\begin{align}
 I(\theta)
 = \left \langle \left( \frac{\partial \log{p(\bm{r}; \theta)}}{\partial \theta} \right) ^2 \right \rangle
 = \left \langle - \frac{\partial^2 \log{p(\bm{r}; \theta)}}{\partial \theta^2} \right \rangle ,
 \label{eq:fisher_info}
\end{align}
where $p(\bm{r}; \theta)$ denotes the probability distribution of the neuronal activity $\bm{r}=(r_1, \ldots, r_{2N})^\top$ responding to the input $\theta$.
A variable transformation from the Gaussian variables to the neural activities gives the explicit form of the distribution function as a convolutional integral of Gaussian functions,
\begin{multline}
 p(\bm{r}; \theta)
 = \frac{1}{(2\pi\sigma_1^2)^{\frac{1}{2}}(2\pi\sigma_0^2)^N}
 \int d\phi \exp \Bigl[ -\frac{1}{2\sigma_1^2}(\phi - \theta)^2 \\
 - \frac{1}{2\sigma_0^2}\sum_{n=1}^N \Bigl(\left(x_n - n^{-\alpha/2}\cos{n\phi}\right)^2 \\
 + \left(y_n - n^{-\alpha/2} \sin{n\phi}\right)^2 \Bigr) \Bigr],
 \label{eq:p}
\end{multline}
where we denote $r_{2n-1}=x_n$ and $r_{2n}=y_n$ for simplicity (See Supplemental material Sec. I for details).

\subsection{Fluctuation-dissipation relationship and the Fisher information of the power-law coding}

To derive an analytical expression for the Fisher information, let us assume that the noise strengths $\sigma_0$ and $\sigma_1$ are sufficiently small so that the probability distribution can be approximated by the multivariate Gaussian distribution,
\begin{align}
    p(\bm{r}; \theta) \approx \frac{1}{\sqrt{(2\pi)^{2N}|\Sigma|}} \exp{\left( - (\bm{r}-\bm{m})^\top \Sigma^{-1}(\bm{r}-\bm{m}) \right)} \label{eq:gauss_approx},
\end{align}
where $\bm{m}$ and $\Sigma$ are the mean and the covariance matrix of the neural activity $\bm{r}$ for given $\theta$, respectively, and $\bm{x}^\top$ denotes the transpose of $\bm{x}$.

\begin{figure}[htb]
  \centering
  \includegraphics[width=8cm]{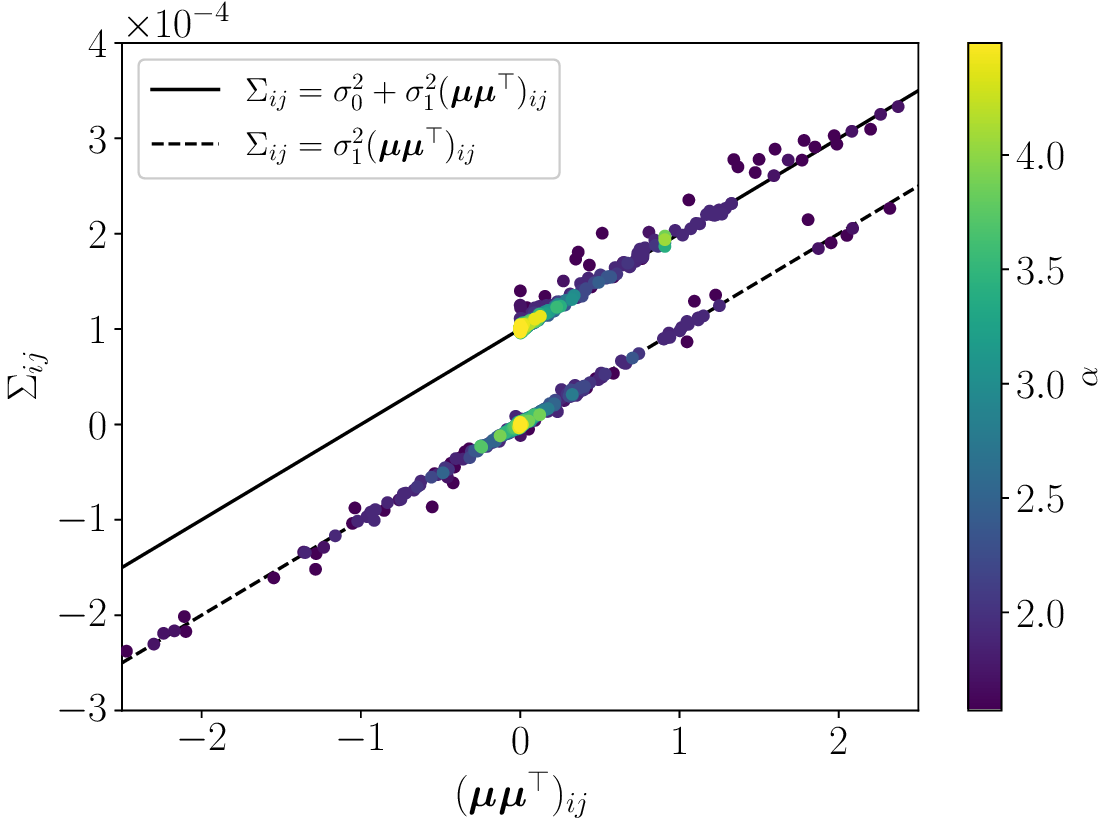}
  \caption{Fluctuation-dissipation relationship of the neural response. Each point represents an element of the covariance matrix $\Sigma_{ij}$ as a function of the corresponding element of the outer product of the susceptibility of the neural response to the input signal $(\bm{\mu} \bm{\mu}^\top)_{ij}$. The color of the points indicates the value of $\alpha$. As predicted by Eq. \eqref{eq:cov_of_gauss_approx}, the covariance is proportional to the outer product for off-diagonal elements. They, however, are shifted by $\sigma_0^2$ to the vertical direction for diagonal elements due to the identity matrix of the first term of the equation. Thick line is $\Sigma_{ij} = \sigma_0^2 + \sigma_1^2 (\bm{\mu}\bm{\mu}^\top)_{ij}$ and the dashed line is $\Sigma_{ij} = \sigma_1^2 (\bm{\mu}\bm{\mu}^\top)_{ij}$. We used $\sigma_1 = 0.01$, $\sigma_0 = 0.01$ and $N=100$ for the plot.}
  \label{fig:fdt}
\end{figure}

Now let us denote the derivative of the mean with respect to the input $\theta$ as $\bm{\mu}=\partial \bm{m}/\partial \theta$, namely, $\bm{\mu}$ is the susceptibility of the mean neural activity to the input signal. Then, we can derive a remarkable relationship between the covariance matrix $\Sigma$ and the susceptibility $\bm{\mu}$:
\begin{align}
 \Sigma = \sigma_0^2 \bm{I} + \sigma_1^2 \bm{\mu} \bm{\mu}^\top ,
 \label{eq:cov_of_gauss_approx}
\end{align}
where $\bm{I}$ is the identity matrix (see Supplemental material Sec. IIA for details). It should be emphasized that the relationship links the fluctuation and the susceptibility of neural activity and thus corresponds to the ``fluctuation-dissipation theorem'' in statistical physics, while the susceptibility appears as a second-order term rather than a linear one \cite{Kubo1966-rg}. To confirm this significant identity we directly measure the covariance matrix and the input susceptibility for the population activity of neurons given by \eqref{eq:coding} for various realizations of the exponent $\alpha$, input noise $\eta$, and neuron noise $\xi$. Figure \ref{fig:fdt} is a scatter plot of elements of the covariance matrix $\Sigma_{ij}$ versus the corresponding outer products of the susceptibility vectors $\left(\bm{\mu} \bm{\mu}^\top\right)_{ij}$. We see that regardless of the parameter realizations both diagonal (points along the solid line) and off-diagonal elements (points along the dashed line) well agrees the theoretical prediction. Thus the relationship surely satisfied for the neural coding.

The above relation also gives the impressive result that the susceptibility is an eigenvector of the covariance matrix, whose eigenvalue is given by using the generalized harmonic function $H_N(x) = \sum_{n=1}^N n^{-x}$, which converges to the Riemann zeta function $\zeta(x) = \sum_{n=1}^{\infty} n^{-x}$ in the limit of large numbers of neurons $N\to\infty$ (see Supplemental material Sec. IIB for details),
\begin{align}
 \begin{split}
  \Sigma \bm{\mu}
  &= \lambda \bm{\mu}
  \\
  \lambda
  &= \sigma_0^2 + \sigma_1^2 H_N(\alpha-2) \\
  &\to \sigma_0^2 + \sigma_1^2 \zeta(\alpha-2). \\
 \end{split}
 \label{eq:eigen}
\end{align}

Putting \eqref{eq:gauss_approx} and \eqref{eq:cov_of_gauss_approx} to \eqref{eq:fisher_info} and using \eqref{eq:eigen}, we have the Fisher information of the power-law coding as
\begin{align}
 I(\theta) = \frac{H_N(\alpha-2)}{\sigma_0^2 + \sigma_1^2 H_N(\alpha-2)},
 \label{eq:I}
\end{align}
which converges to
\begin{align}
 I_1(\alpha) := \frac{\zeta(\alpha - 2)}{\sigma_0^2 + \sigma_1^2\zeta(\alpha - 2)} ,
 \label{eq:I1}
\end{align}
in the limit of large number of neurons.

The generalization to $D>1$, where neurons encodes a higher dimensional input, is almost straightforward (see Supplemental material Sec. III for details). The Fisher information for multivariate input signal $\bm{\theta}$ is defined in the matrix form,
\begin{align*}
 I_{ij}(\bm{\theta})
 = \left\langle - \frac{\partial^2 \log p(\bm{r}; \bm{\theta})}{\partial \theta_i \partial \theta_j} \right\rangle .
\end{align*}
If the Gaussian noise with strength $\sigma_i$ independently affects to the $i$th input component $\theta_i$, then using the same argument as above, we can show that the Fisher information matrix is diagonal and, in the limit of large numbers of neurons, the $i$th diagonal component, which evaluates the ability of the power-law coding for the $i$th input signal $\theta_i$, converges to
\begin{align}
 I_{ii}(\bm{\theta}) \to
 I_D(\alpha) :=
 \frac{\zeta{\left(\alpha-2/D\right)}}
 {\sigma_0^2 D V_D^{2/D} / 4 + \sigma_i^2 \zeta{\left(\alpha-2/D\right)}},
 \label{eq:Id}
\end{align}
where $V_D = \pi^{D/2}/\Gamma(D/2 + 1) $ is the volume of the unit $D$-ball. This reproduces \eqref{eq:I1} as a special case and, in the limit of large input dimension, $D\to\infty$, will converges to
\begin{align}
    I_\infty(\alpha) = \frac{\zeta(\alpha)}{e \pi \sigma_0^2 / 2 + \sigma_i^2 \zeta(\alpha)}
\end{align}
because $\Gamma\left(z+1\right)\approx \sqrt{2\pi z} (z/e)^z$ for large $z$.

\begin{figure}[htb]
  \centering
  \includegraphics[width=8cm]{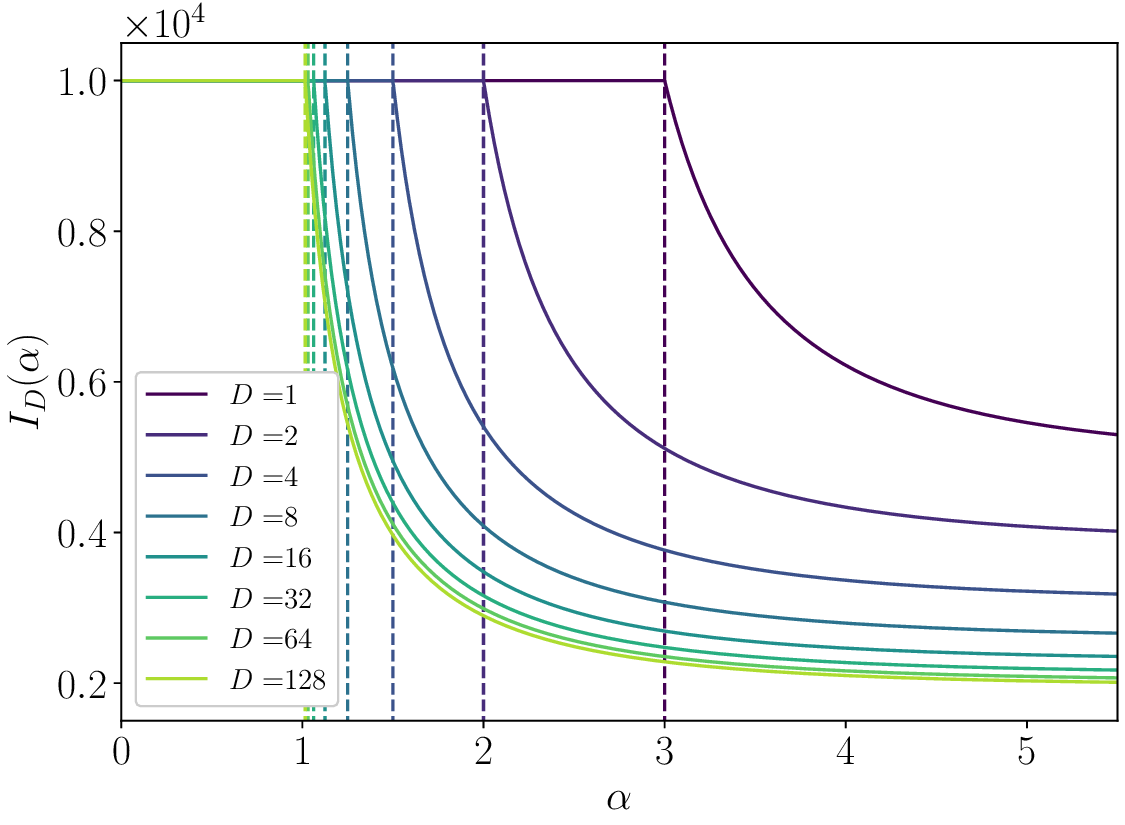}
  \caption{Analytically derived Fisher information $I_D$ as a function of the power-law exponent $\alpha$ for various values of the input dimension $D$. Vertical dashed lines indicate the critical values of the exponent $\alpha_c=1+2/D$. For all $D$, the information rapidly decreases as $\alpha$ increases for $\alpha>\alpha_c$. However, counterintuitively, they keep a constant value for $\alpha<\alpha_c$ even though the neural representation is intrinsically non-differentiable, the neural manifold is fractal, and thus, the population response is highly sensitive to the perturbation in the regime. Here, we used $\sigma_0=\sigma_1=0.01$.}
  \label{fig:fisher_info_and_d}
\end{figure}

Figure \ref{fig:fisher_info_and_d} shows the derived analytical expression of the Fisher information \eqref{eq:Id} as functions of the power-law exponent $\alpha$ for various values of the input dimension $D$. One can see three significant features from the plots. First, the critical exponent $\alpha_c=1+2/D$ been reported in the previous study actually gives the transition point in the sense that the derivative of the Fisher information with respect to the exponent $\alpha$ is discontinuous only at $\alpha=\alpha_c$. Second, for $\alpha>\alpha_c$, each Fisher information is monotonically decreases with increasing $\alpha$, indicating that when the slope of the eigenspectrum of the covariance matrix is steeper than the critical slope, the coding performance deteriorates as the slope increases, which is natural because the steeper the slope the less neural activity is used in the coding.

However, the third and counterintuitive point is that the Fisher information is kept constant and the coding does not degrade its performance for $\alpha<\alpha_c$ where the slope of the power-law decay is gentler than the critical slope, and thus, the map from the input space to the neural space defined by the coding \eqref{eq:coding} is non-differentiable and fractal even in the absence of noise. This result is different from what was previously thought. The Riemann zeta function $\zeta(\alpha-2/D)$ in \eqref{eq:Id} does indeed diverge to infinity when $\alpha \leq \alpha_c$ because the zeta function $\zeta(x)$ diverges to infinity for $x \leq 1$. However, this does not imply either the divergence or the vanishment of the Fisher information. Rather, the information is held constant at $1/\sigma_1^2$, regardless of the neural noise strength $\sigma_0$. Thus, contrary to the previous conjecture, the coding performance is not directly related to the differentiability of the map defined by the coding.

Why does the coding not degrade its performance even in the chaotic regime where the code is extremely sensitive to the perturbation? As shown in the previous work, the map of the coding is indeed non-differentiable for $\alpha \leq \alpha_c$. However, this non-differentiability, induced by the decrease of the power-law slope, is due to the increase of the variance of the eigenspectrum of the components with larger $n$, i.e. higher frequencies, and importantly, the activities of the lower frequency components relatively remain intact. Thus, by using a decoder that appropriately focuses on the directions of the low-frequency components, one can still decode enough information even from the fractal neural coding, which means that the neural activity still contains enough information of the input signals and its coding performance remains intact.

\subsection{Energy-information tradeoff optimizes the cortical critical power law}

A natural question then is why the brain uses the critical exponent $\alpha_c$ rather than one of the other smaller values of $\alpha$ that can also achieve the best coding performance. So far, we have not considered metabolic, or energetic, costs required for the neural activities \cite{Laughlin1998-yu,Lennie2003-td}. However, in fact, the smaller $\alpha$ implies that the larger amount of neural activity is evoked to represent the input signals, which may explain the optimality of the exponent $\alpha_c$ for the brain's encoding.

To see that this is indeed the case, we introduce a performance measure of the signal encoding with including an energetic cost of neural activity. Whereas the biologically realistic description of the metabolic cost of neural activity is beyond the scope of the present work, a natural definition of this will be the sum of the mean square of the response activities of all neurons over the input stimuli and the noise, which converges to the zeta function in the limit of large numbers of neurons $N\to\infty$ in the present setting:
\begin{align*}
 \sum_{j=1}^{N} \frac{1}{2\pi}\int_0^{2\pi} \langle r_i\left(\theta\right) \rangle_{\xi_i, \eta}^2 d\theta
 = \sum_{n=1}^{N} n^{-\alpha}
 \to \zeta\left(\alpha\right) .
\end{align*}
Therefore, the sum of the Fisher information and the energy cost gives an energy-aware performance measure of the power-law coding
\begin{align}
    J_D(\alpha) = I_D(\theta) - \gamma \zeta(\alpha), \label{eq:energy-aware_fisher_info}
\end{align}
where the regularization parameter must satisfy $\gamma > 0$ because lower energy consumption is desirable.

\begin{figure}[htb]
  \centering
  \includegraphics[width=8cm]{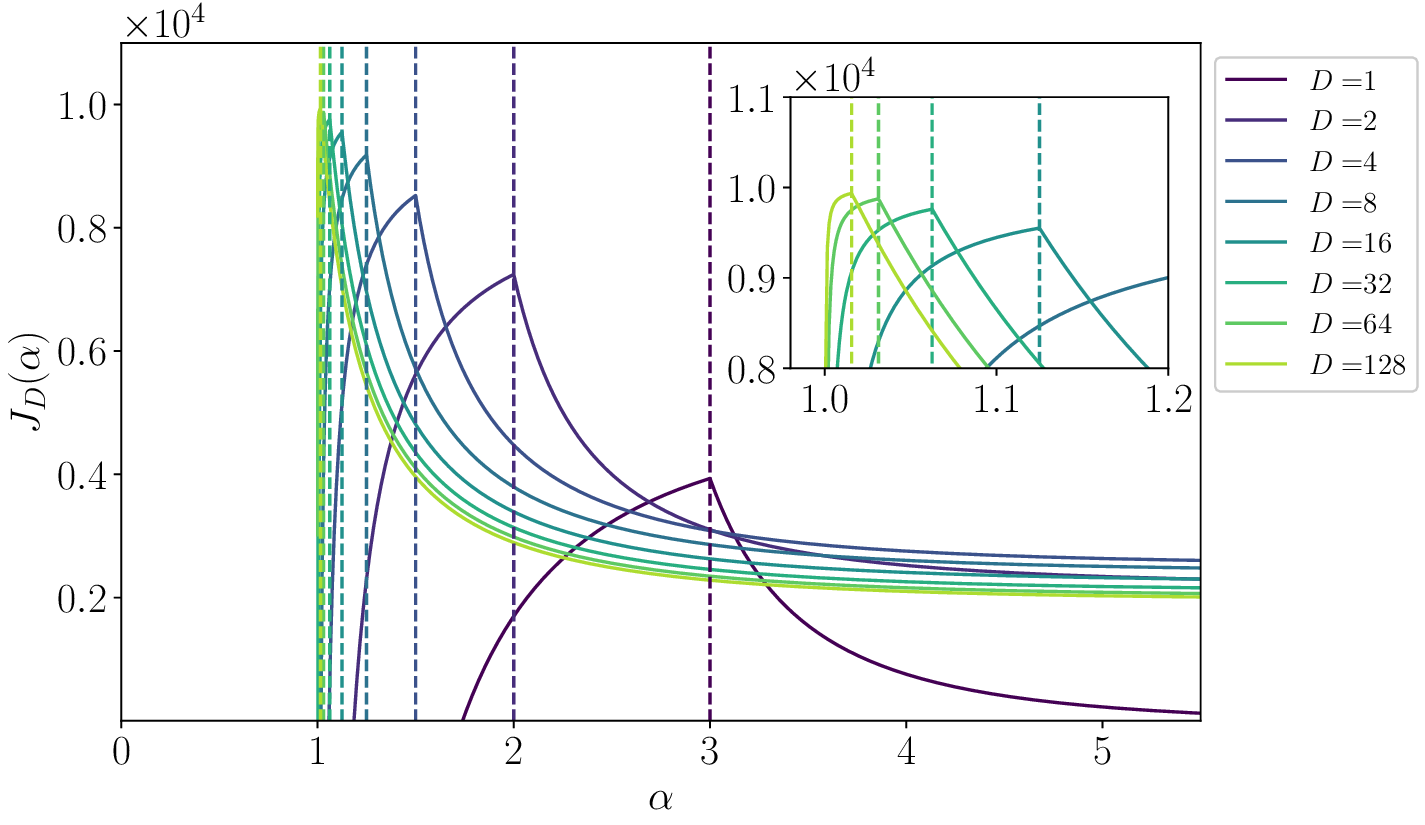}
  \caption{The energy-aware performance measure of the power-law coding $J_D$ as a function of the power-law exponent $\alpha$ for various values of the input dimension $D$. Vertical dashed lines indicate the critical values of the exponent $\alpha_c=1+2/D$. Inset represents the magnification of the plot around $\alpha=1$. Due to the energetic cost term, different from Fig. \ref{fig:fisher_info_and_d}, regardless of input dimension, all the lines take their optimal value exactly where the power-law exponent is critical $\alpha=\alpha_c$. Please also note that the critical value $\alpha_c$ seems almost the exclusive choice of the exponent for large input dimension $D$ as indicated by the delta-function-like curve. Here, $\sigma_i =\sigma_0 = 0.01$, and sufficiently small values of $\gamma$ that do not overwhelm the first term of Eq. \eqref{eq:energy-aware_fisher_info} are used. (see Supplementary material IV for details).}
  \label{fig:fisher_info_and_energy_constraint}
\end{figure}

Figure \ref{fig:fisher_info_and_energy_constraint} shows the energy-aware performance Eq. \eqref{eq:energy-aware_fisher_info} as functions of the exponent $\alpha$ for various values of the input dimension $D$. In contrast to Fig. \ref{fig:fisher_info_and_d}, the performances are maximized at the critical exponent $\alpha=\alpha_c$, regardless of the input dimension. This result is natural because the first term of the energy-aware performance is flat for $\alpha<\alpha_c$ as shown in Fig. \ref{fig:fisher_info_and_d}, while the second term of the energy cost given by the negative zeta function is a monotonically increasing with $\alpha$. In other words, while we have used arbitrarily small values of $\lambda$ for the plot, the result is robust to the actual choice of values of the regularization coefficient unless it is too large to overwhelm the first term (see Supplemental material for details). The same argument also implies that the energy term need not be the sum of the square of the neural activity in order to give the same optimality of the critical exponent, as long as the term is the monotonically increasing function of $\alpha$. Thus, we can conclude that, for a broad class of energy terms, the trade-off between Fisher information and energy consumption explains the optimality of the critical exponent observed in the brain.

\subsection{Maximum likelihood estimator for the power-law code}

\begin{figure*}[htb]
\begin{center}
\includegraphics[width=18cm]{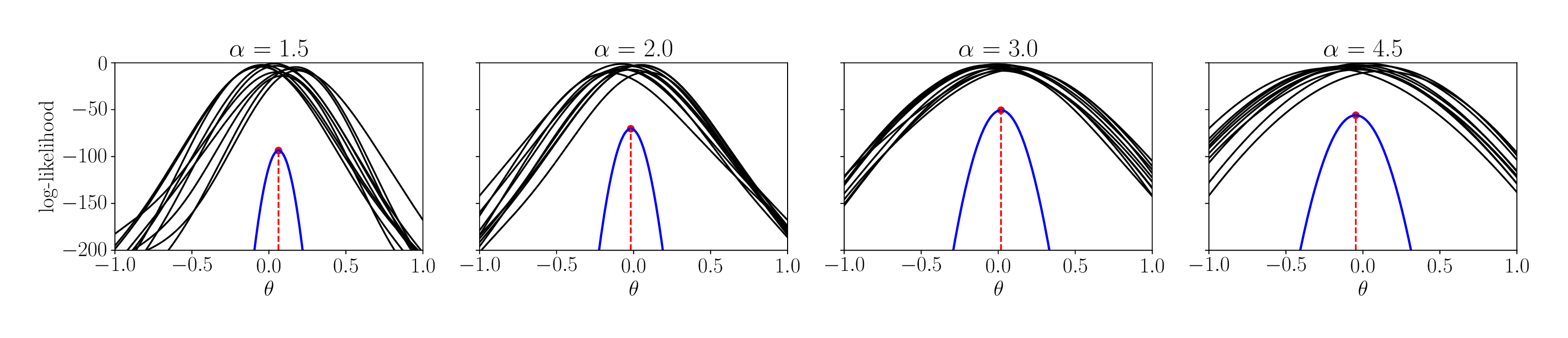}
\caption{Outline of the maximum likelihood estimation of the input stimulus from the power-law population coding. The logarithm of the posterior functions $\log P(r_i|\theta)$ (black lines) are directly calculated by numerical integration of the posterior distribution \eqref{eq:p} for various realizations of neural activities \eqref{eq:coding} as functions of the variable $\theta$. Each panel exemplifies the functions for values of exponent $\alpha$, where we show just ten realizations of the posterior functions to illustrate the procedure. The maximum likelihood estimate $\hat{\theta}$ (the red point and the red dashed line) of the input signal is given as the value of $\theta$ that maximizes the log-likelihood function \eqref{eq:loglikelihood} (the blue line) that is the sum of the logarithm of the posterior functions.}
\label{fig:mll}
\end{center}
\end{figure*}

To validate the theoretical predictions, we actually construct the maximum likelihood decoder for the power-law coding and directly measure the variance of the estimation errors of the input stimulus $\theta$ to compare it with the inverse of the predicted Fisher information. For a given set of $M$ observations of neural activities $\bm{r}^{(m)}(m=1,\ldots,M)$ responding to the input stimulus $\theta$, the log-likelihood function of the input stimulus is given by
\begin{align}
 L\left(\theta;\bm{r}^{(1)},\ldots,\bm{r}^{(M)}\right)=\sum_{m=1}^M\log{p(\bm{r}^{(m)}; \theta}), \label{eq:loglikelihood}
\end{align}
where, $p$ is the probability density function given by \eqref{eq:p}. We prepare a set of $M$ realizations of the activity of $N$ neurons based on \eqref{eq:coding}, then put them into the log-likelihood function \eqref{eq:loglikelihood} and numerically integrate probability distributions \eqref{eq:p} for values of $\theta$ to find the maximum likelihood estimate for the neural activity, i.e., the value of $\theta$ that maximizes the log-likelihood function (Fig. \ref{fig:mll}). By repeating the procedure, we compute the inverse of the variance of the estimates, which must asymptotically converge to the Fisher information in the limit of large numbers of observations $M\to\infty$ due to the Cramer-Rao theorem.

\begin{figure}[htb]
\begin{center}
\includegraphics[width=8cm,bb=0 0 512 403]{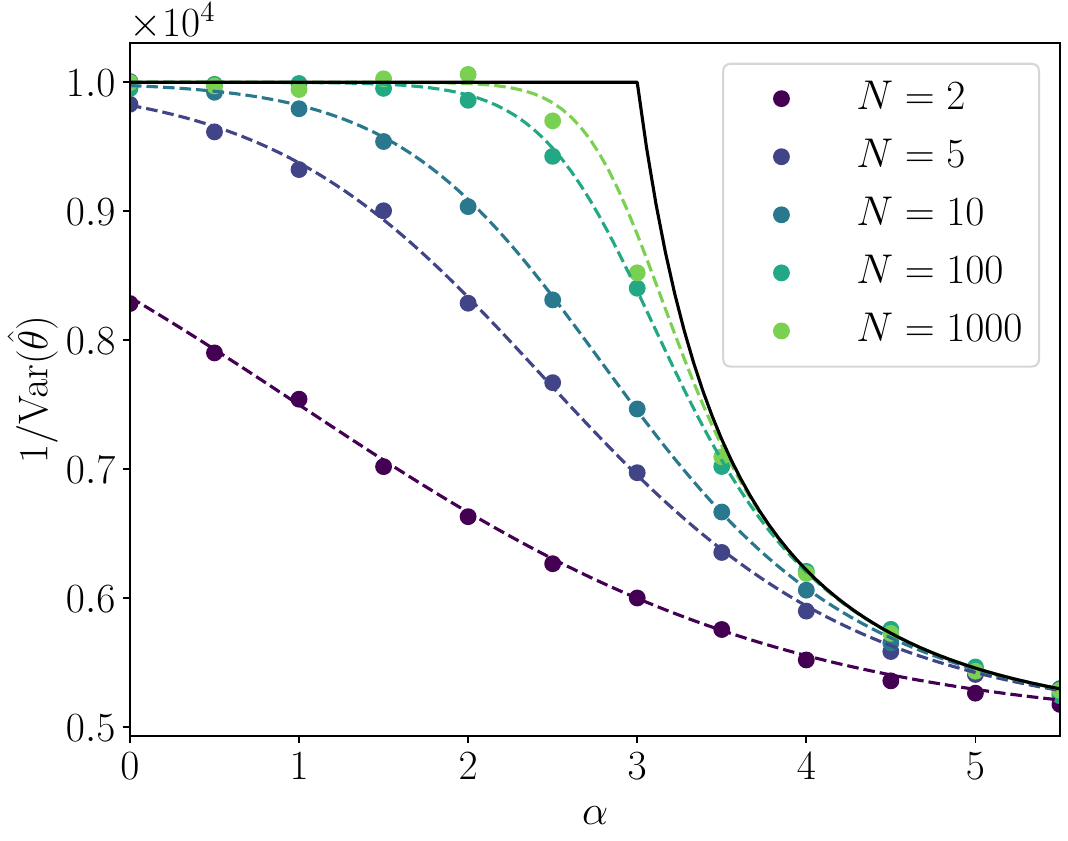}
\caption{The inverse of the variance of the estimation error of the maximum likelihood estimation of the input stimulus for the power-law population coding. This inverse will be the asymptotical equivalence of the Fisher information of the coding. Using direct numerical integration of the posterior distribution, we obtained the variance form $M = 10^5$ realizations of the numerical estimation for various values of exponent $\alpha$ and the numbers of neurons $N$ (colored circle points). Dashed lines are the theoretical prediction corresponding to the finite size Fisher information \eqref{eq:I}, and the solid line is the one in the limit of large numbers of neurons ($N \rightarrow \infty$) that is given Eq. \eqref{eq:I1}.}
\label{fig:f3}
\end{center}
\end{figure}

Figure \ref{fig:f3} shows the numerically obtained inverse of the variance of the estimation error as functions of the power-law exponent $\alpha$ for various values of $N$. Each dotted line corresponds to the derived Fisher information, Eq. \eqref{eq:I}. One can confirm that numerical results well agree with the analytical predictions, and they monotonically converge to the solid line representing the Fisher information in the limit of large number of neurons, Eq. \eqref{eq:I1}.

\section{Discussion}

To provide a quantitative evaluation of the coding ability of the power-law stimulus representation of the population of cortical neurons, we have developed an analytically tractable model of neural coding whose spectrum of the covariance matrix follows the power law observed experimentally in the cortex. Owing to the remarkable relationship connecting the variance and the susceptibility of neural responses to the input stimuli, that is, a kind of fluctuation-dissipation theorem of neural responses, the theory allows us to explicitly derive the Fisher information of the power-law coding. Contrary to previous conjectures, the derived Fisher information demonstrates that neural coding ability is not degraded even in non-differentiable and fractal neural manifolds, where neural responses must be largely perturbed by noise included in input signals and neural activities. Conversely, our findings indicate that the introduction of a minimal metabolic cost of neural activities robustly makes the critical power-law neural response the optimal neural coding. Therefore, the experimentally observed power-law exponent gives the best balance between the energetic cost and the Fisher information of the neural coding. Consequently, the critical response observed in the brain does not merely achieve a balance between the expressivity and the robustness of the coding; it additionally strikes a balance between these two factors and the energetic cost, which must be a crucial factor for biological computation in the real world.

In this paper, assuming the visual cortex that the power-law response has been experimentally confirmed, we have restricted ourselves to the case where the input stimuli to the neurons are expressed by Fourier basis functions. However, as shown by the fluctuation-dissipation relationship, the basic framework developed cannot be restricted to the Fourier basis receptive field, whereas the actual expression of the Fisher information and the details of the tradeoff between energetic and information costs may differ from the result of the paper. For instance, introducing the Gabor filter-type receptive field, which is closer to the actual realization of the biological realization than the simple Fourier receptive field discussed here, may give an additional factor in the representation of the Fisher information. Moreover, since stimulus responses and receptive fields are generally different across cortical areas responsible for different modalities, these varieties may make different power-law exponents the critical one that realizes the best balance of energy-info tradeoff. Because recent experiments started to report that the experimentally observed power-law exponent slightly differs across areas of the cortex and across species \cite{Kong2022-fh,Morales2023-ii}, it would be an important future task to apply the theory to stimulus representation with various receptive fields used in cortical regions for other modalities to reveal relationships between the observed exponents and characteristic features, such as data structure of input stimuli, across modalities.

We have introduced the power-law dependency of neural activities, a priori, here and assumed that each neuron independently responds to its stimulus with the power-law amplitude. However, it is naturally expected that this power-law dependency itself must be realized in a network of neurons that is activated by external or internal signals with some correlation. Pioneering studies have revealed that the Fisher information is generally influenced by the activity correlation \cite{Sompolinsky2001-dl,Moreno-Bote2014-dh} whereas its effect is not very large when the correlation is originated from recurrent neural activities \cite{Moreno-Bote2014-dh}. Correlation of neurons can also be very important here because it must directly be related to the power-law responses because the power-law appears in eigenspectral of the covariance matrix of response neural activities. While it remains a possibility that the covariance is purely induced by the structure of the input connections to the encoder neurons, it seems more natural that the covariance, and thus the power-law itself, is originated form interaction of the input and recurrent connections among encoder neurons \cite{Hu2022-ib,Huang2022-sa,Wardak2022-xu}. Developing analysis of the power-law coding with combining theories of neural correlation of recurrent networks is thus important. Temporal structure of neural responses such as transient neural activities, for instance, may play an important role in these cases \cite{Vyas2020-eu,Dubreuil2022-qt}. It is also interesting remaining work to confirm the theoretical predictions by using network of realistic neuron models.

The theory developed here bears a striking resemblance to a framework of the Shannon-Hartley theorem of information theory describing the capacity of an analog communication channel under the limitation of signal power \cite{cover1999elements}. The theory states that the channel capacity, i.e., the upper bound of the information rate of the data can be transmitted through the channel without error, is given as the function of the ratio of the strengths of input signals and the additive white Gaussian noise. One can easily see the similarity between the communication using the noisy channel and the flow of the stimulus representation, that is, from the encoding of input stimulus to the activity of the population of neurons to the decoding of the value from the neural activity (Fig. \ref{fig:fig1}a). Studies to find closer relationships between them must be an interesting future direction.

Because the Principal Component Analysis is based on a covariance matrix of neural activities, it can give only a linear structure of the neural manifold, whereas the manifold can generally have a higher-order structure \cite{De2023-su}. Thus, geometric indices such as curvature of the neural manifold may play an important role in the optimal stimulus encoding in addition to the power-law exponent of the eigenspectral of the covariance matrix. Because the Fisher information matrix works as a suitable metric of statistical manifolds of probability distributions, it will also be an interesting future direction to extend the theory of optimal coding on the neural manifolds to include higher-order structures considering the information geometry \cite{amari2000methods}.

Another important critical phenomenon observed for populations of cortical neurons is neural avalanche, in which the size and the duration of spontaneously generated neural activity follow the power-law distribution \cite{Beggs2003-ac,De_Arcangelis2006-wx,Fosque2021-fw}. Along with the power-law coding discussed in this study, the power-law distribution of the neural avalanche also gives strong evidence that actual cortical networks work in the so-called edge-of-chaos region, which is near the critical transition point of the stability of neural responses. While the relationship between the criticality of the neural avalanche and that of the power-law coding has not been clarified yet, probably because the former considers only spontaneous activities while the latter does only stimulus-evoked activities, we can naturally expect that these two critical phenomena have a tight relationship because both of them essentially refer critical correlation structure of neural activities. As mentioned in the above section, extending the theory developed here to recurrent networks of neurons might help to reveal the hidden relation between these two neural criticalities.

\section*{Acknowledgements}

T. T. discloses support for the research of this work by JST SPRING Grant Number JPMJSP2110. J. T. discloses support for publication of this work by JSPS KAKENHI Grant Number 24K15104 and 23K21352, and Japan Agency for Medical Research and Development (AMED) AMED-CREST 23gm1510005h0003.

\bibliography{refs}

\begin{thebibliography}{49}%
\makeatletter
\providecommand \@ifxundefined [1]{%
 \@ifx{#1\undefined}
}%
\providecommand \@ifnum [1]{%
 \ifnum #1\expandafter \@firstoftwo
 \else \expandafter \@secondoftwo
 \fi
}%
\providecommand \@ifx [1]{%
 \ifx #1\expandafter \@firstoftwo
 \else \expandafter \@secondoftwo
 \fi
}%
\providecommand \natexlab [1]{#1}%
\providecommand \enquote  [1]{``#1''}%
\providecommand \bibnamefont  [1]{#1}%
\providecommand \bibfnamefont [1]{#1}%
\providecommand \citenamefont [1]{#1}%
\providecommand \href@noop [0]{\@secondoftwo}%
\providecommand \href [0]{\begingroup \@sanitize@url \@href}%
\providecommand \@href[1]{\@@startlink{#1}\@@href}%
\providecommand \@@href[1]{\endgroup#1\@@endlink}%
\providecommand \@sanitize@url [0]{\catcode `\\12\catcode `\$12\catcode
  `\&12\catcode `\#12\catcode `\^12\catcode `\_12\catcode `\%12\relax}%
\providecommand \@@startlink[1]{}%
\providecommand \@@endlink[0]{}%
\providecommand \url  [0]{\begingroup\@sanitize@url \@url }%
\providecommand \@url [1]{\endgroup\@href {#1}{\urlprefix }}%
\providecommand \urlprefix  [0]{URL }%
\providecommand \Eprint [0]{\href }%
\providecommand \doibase [0]{https://doi.org/}%
\providecommand \selectlanguage [0]{\@gobble}%
\providecommand \bibinfo  [0]{\@secondoftwo}%
\providecommand \bibfield  [0]{\@secondoftwo}%
\providecommand \translation [1]{[#1]}%
\providecommand \BibitemOpen [0]{}%
\providecommand \bibitemStop [0]{}%
\providecommand \bibitemNoStop [0]{.\EOS\space}%
\providecommand \EOS [0]{\spacefactor3000\relax}%
\providecommand \BibitemShut  [1]{\csname bibitem#1\endcsname}%
\let\auto@bib@innerbib\@empty
\bibitem [{\citenamefont {Barlow}\ \emph {et~al.}(1961)\citenamefont {Barlow}
  \emph {et~al.}}]{barlow1961possible}%
  \BibitemOpen
  \bibfield  {author} {\bibinfo {author} {\bibfnamefont {H.~B.}\ \bibnamefont
  {Barlow}} \emph {et~al.},\ }\bibfield  {title} {\bibinfo {title} {Possible
  principles underlying the transformation of sensory messages},\ }\href@noop
  {} {\bibfield  {journal} {\bibinfo  {journal} {Sensory communication}\
  }\textbf {\bibinfo {volume} {1}},\ \bibinfo {pages} {217} (\bibinfo {year}
  {1961})}\BibitemShut {NoStop}%
\bibitem [{\citenamefont {Simoncelli}\ and\ \citenamefont
  {Olshausen}(2001)}]{Simoncelli2001-ar}%
  \BibitemOpen
  \bibfield  {author} {\bibinfo {author} {\bibfnamefont {E.~P.}\ \bibnamefont
  {Simoncelli}}\ and\ \bibinfo {author} {\bibfnamefont {B.~A.}\ \bibnamefont
  {Olshausen}},\ }\bibfield  {title} {\bibinfo {title} {Natural image
  statistics and neural representation},\ }\href
  {https://doi.org/10.1146/annurev.neuro.24.1.1193} {\bibfield  {journal}
  {\bibinfo  {journal} {Annu. Rev. Neurosci.}\ }\textbf {\bibinfo {volume}
  {24}},\ \bibinfo {pages} {1193} (\bibinfo {year} {2001})}\BibitemShut
  {NoStop}%
\bibitem [{\citenamefont {Churchland}\ \emph {et~al.}(2012)\citenamefont
  {Churchland}, \citenamefont {Cunningham}, \citenamefont {Kaufman},
  \citenamefont {Foster}, \citenamefont {Nuyujukian}, \citenamefont {Ryu},\
  and\ \citenamefont {Shenoy}}]{Churchland2012-iu}%
  \BibitemOpen
  \bibfield  {author} {\bibinfo {author} {\bibfnamefont {M.~M.}\ \bibnamefont
  {Churchland}}, \bibinfo {author} {\bibfnamefont {J.~P.}\ \bibnamefont
  {Cunningham}}, \bibinfo {author} {\bibfnamefont {M.~T.}\ \bibnamefont
  {Kaufman}}, \bibinfo {author} {\bibfnamefont {J.~D.}\ \bibnamefont {Foster}},
  \bibinfo {author} {\bibfnamefont {P.}~\bibnamefont {Nuyujukian}}, \bibinfo
  {author} {\bibfnamefont {S.~I.}\ \bibnamefont {Ryu}},\ and\ \bibinfo {author}
  {\bibfnamefont {K.~V.}\ \bibnamefont {Shenoy}},\ }\bibfield  {title}
  {\bibinfo {title} {Neural population dynamics during reaching},\ }\href
  {https://doi.org/10.1038/nature11129} {\bibfield  {journal} {\bibinfo
  {journal} {Nature}\ }\textbf {\bibinfo {volume} {487}},\ \bibinfo {pages}
  {51} (\bibinfo {year} {2012})}\BibitemShut {NoStop}%
\bibitem [{\citenamefont {Rigotti}\ \emph {et~al.}(2013)\citenamefont
  {Rigotti}, \citenamefont {Barak}, \citenamefont {Warden}, \citenamefont
  {Wang}, \citenamefont {Daw}, \citenamefont {Miller},\ and\ \citenamefont
  {Fusi}}]{Rigotti2013-oy}%
  \BibitemOpen
  \bibfield  {author} {\bibinfo {author} {\bibfnamefont {M.}~\bibnamefont
  {Rigotti}}, \bibinfo {author} {\bibfnamefont {O.}~\bibnamefont {Barak}},
  \bibinfo {author} {\bibfnamefont {M.~R.}\ \bibnamefont {Warden}}, \bibinfo
  {author} {\bibfnamefont {X.-J.}\ \bibnamefont {Wang}}, \bibinfo {author}
  {\bibfnamefont {N.~D.}\ \bibnamefont {Daw}}, \bibinfo {author} {\bibfnamefont
  {E.~K.}\ \bibnamefont {Miller}},\ and\ \bibinfo {author} {\bibfnamefont
  {S.}~\bibnamefont {Fusi}},\ }\bibfield  {title} {\bibinfo {title} {The
  importance of mixed selectivity in complex cognitive tasks},\ }\href
  {https://doi.org/10.1038/nature12160} {\bibfield  {journal} {\bibinfo
  {journal} {Nature}\ }\textbf {\bibinfo {volume} {497}},\ \bibinfo {pages}
  {585} (\bibinfo {year} {2013})}\BibitemShut {NoStop}%
\bibitem [{\citenamefont {Sadtler}\ \emph {et~al.}(2014)\citenamefont
  {Sadtler}, \citenamefont {Quick}, \citenamefont {Golub}, \citenamefont
  {Chase}, \citenamefont {Ryu}, \citenamefont {Tyler-Kabara}, \citenamefont
  {Yu},\ and\ \citenamefont {Batista}}]{Sadtler2014-uf}%
  \BibitemOpen
  \bibfield  {author} {\bibinfo {author} {\bibfnamefont {P.~T.}\ \bibnamefont
  {Sadtler}}, \bibinfo {author} {\bibfnamefont {K.~M.}\ \bibnamefont {Quick}},
  \bibinfo {author} {\bibfnamefont {M.~D.}\ \bibnamefont {Golub}}, \bibinfo
  {author} {\bibfnamefont {S.~M.}\ \bibnamefont {Chase}}, \bibinfo {author}
  {\bibfnamefont {S.~I.}\ \bibnamefont {Ryu}}, \bibinfo {author} {\bibfnamefont
  {E.~C.}\ \bibnamefont {Tyler-Kabara}}, \bibinfo {author} {\bibfnamefont
  {B.~M.}\ \bibnamefont {Yu}},\ and\ \bibinfo {author} {\bibfnamefont {A.~P.}\
  \bibnamefont {Batista}},\ }\bibfield  {title} {\bibinfo {title} {Neural
  constraints on learning},\ }\href {https://doi.org/10.1038/nature13665}
  {\bibfield  {journal} {\bibinfo  {journal} {Nature}\ }\textbf {\bibinfo
  {volume} {512}},\ \bibinfo {pages} {423} (\bibinfo {year}
  {2014})}\BibitemShut {NoStop}%
\bibitem [{\citenamefont {Chung}\ \emph {et~al.}(2018)\citenamefont {Chung},
  \citenamefont {Lee},\ and\ \citenamefont {Sompolinsky}}]{Chung2018-og}%
  \BibitemOpen
  \bibfield  {author} {\bibinfo {author} {\bibfnamefont {S.}~\bibnamefont
  {Chung}}, \bibinfo {author} {\bibfnamefont {D.~D.}\ \bibnamefont {Lee}},\
  and\ \bibinfo {author} {\bibfnamefont {H.}~\bibnamefont {Sompolinsky}},\
  }\bibfield  {title} {\bibinfo {title} {Classification and geometry of general
  perceptual manifolds},\ }\href {https://doi.org/10.1103/PhysRevX.8.031003}
  {\bibfield  {journal} {\bibinfo  {journal} {Phys. Rev. X}\ }\textbf {\bibinfo
  {volume} {8}},\ \bibinfo {pages} {031003} (\bibinfo {year}
  {2018})}\BibitemShut {NoStop}%
\bibitem [{\citenamefont {Gallego}\ \emph {et~al.}(2018)\citenamefont
  {Gallego}, \citenamefont {Perich}, \citenamefont {Naufel}, \citenamefont
  {Ethier}, \citenamefont {Solla},\ and\ \citenamefont
  {Miller}}]{Gallego2018-qq}%
  \BibitemOpen
  \bibfield  {author} {\bibinfo {author} {\bibfnamefont {J.~A.}\ \bibnamefont
  {Gallego}}, \bibinfo {author} {\bibfnamefont {M.~G.}\ \bibnamefont {Perich}},
  \bibinfo {author} {\bibfnamefont {S.~N.}\ \bibnamefont {Naufel}}, \bibinfo
  {author} {\bibfnamefont {C.}~\bibnamefont {Ethier}}, \bibinfo {author}
  {\bibfnamefont {S.~A.}\ \bibnamefont {Solla}},\ and\ \bibinfo {author}
  {\bibfnamefont {L.~E.}\ \bibnamefont {Miller}},\ }\bibfield  {title}
  {\bibinfo {title} {Cortical population activity within a preserved neural
  manifold underlies multiple motor behaviors},\ }\href
  {https://doi.org/10.1038/s41467-018-06560-z} {\bibfield  {journal} {\bibinfo
  {journal} {Nat. Commun.}\ }\textbf {\bibinfo {volume} {9}},\ \bibinfo {pages}
  {4233} (\bibinfo {year} {2018})}\BibitemShut {NoStop}%
\bibitem [{\citenamefont {Stringer}\ \emph {et~al.}(2019)\citenamefont
  {Stringer}, \citenamefont {Pachitariu}, \citenamefont {Steinmetz},
  \citenamefont {Carandini},\ and\ \citenamefont {Harris}}]{Stringer2019-dw}%
  \BibitemOpen
  \bibfield  {author} {\bibinfo {author} {\bibfnamefont {C.}~\bibnamefont
  {Stringer}}, \bibinfo {author} {\bibfnamefont {M.}~\bibnamefont
  {Pachitariu}}, \bibinfo {author} {\bibfnamefont {N.}~\bibnamefont
  {Steinmetz}}, \bibinfo {author} {\bibfnamefont {M.}~\bibnamefont
  {Carandini}},\ and\ \bibinfo {author} {\bibfnamefont {K.~D.}\ \bibnamefont
  {Harris}},\ }\bibfield  {title} {\bibinfo {title} {High-dimensional geometry
  of population responses in visual cortex},\ }\href
  {https://doi.org/10.1038/s41586-019-1346-5} {\bibfield  {journal} {\bibinfo
  {journal} {Nature}\ }\textbf {\bibinfo {volume} {571}},\ \bibinfo {pages}
  {361} (\bibinfo {year} {2019})}\BibitemShut {NoStop}%
\bibitem [{\citenamefont {Gallego}\ \emph {et~al.}(2020)\citenamefont
  {Gallego}, \citenamefont {Perich}, \citenamefont {Chowdhury}, \citenamefont
  {Solla},\ and\ \citenamefont {Miller}}]{Gallego2020-ap}%
  \BibitemOpen
  \bibfield  {author} {\bibinfo {author} {\bibfnamefont {J.~A.}\ \bibnamefont
  {Gallego}}, \bibinfo {author} {\bibfnamefont {M.~G.}\ \bibnamefont {Perich}},
  \bibinfo {author} {\bibfnamefont {R.~H.}\ \bibnamefont {Chowdhury}}, \bibinfo
  {author} {\bibfnamefont {S.~A.}\ \bibnamefont {Solla}},\ and\ \bibinfo
  {author} {\bibfnamefont {L.~E.}\ \bibnamefont {Miller}},\ }\bibfield  {title}
  {\bibinfo {title} {Long-term stability of cortical population dynamics
  underlying consistent behavior},\ }\href
  {https://doi.org/10.1038/s41593-019-0555-4} {\bibfield  {journal} {\bibinfo
  {journal} {Nat. Neurosci.}\ }\textbf {\bibinfo {volume} {23}},\ \bibinfo
  {pages} {260} (\bibinfo {year} {2020})}\BibitemShut {NoStop}%
\bibitem [{\citenamefont {Yoshida}\ and\ \citenamefont
  {Ohki}(2020)}]{Yoshida2020-ji}%
  \BibitemOpen
  \bibfield  {author} {\bibinfo {author} {\bibfnamefont {T.}~\bibnamefont
  {Yoshida}}\ and\ \bibinfo {author} {\bibfnamefont {K.}~\bibnamefont {Ohki}},\
  }\bibfield  {title} {\bibinfo {title} {Natural images are reliably
  represented by sparse and variable populations of neurons in visual cortex},\
  }\href {https://doi.org/10.1038/s41467-020-14645-x} {\bibfield  {journal}
  {\bibinfo  {journal} {Nat. Commun.}\ }\textbf {\bibinfo {volume} {11}},\
  \bibinfo {pages} {872} (\bibinfo {year} {2020})}\BibitemShut {NoStop}%
\bibitem [{\citenamefont {Chung}\ and\ \citenamefont
  {Abbott}(2021)}]{Chung2021-ao}%
  \BibitemOpen
  \bibfield  {author} {\bibinfo {author} {\bibfnamefont {S.}~\bibnamefont
  {Chung}}\ and\ \bibinfo {author} {\bibfnamefont {L.~F.}\ \bibnamefont
  {Abbott}},\ }\bibfield  {title} {\bibinfo {title} {Neural population
  geometry: An approach for understanding biological and artificial neural
  networks},\ }\href {https://doi.org/10.1016/j.conb.2021.10.010} {\bibfield
  {journal} {\bibinfo  {journal} {Curr. Opin. Neurobiol.}\ }\textbf {\bibinfo
  {volume} {70}},\ \bibinfo {pages} {137} (\bibinfo {year} {2021})}\BibitemShut
  {NoStop}%
\bibitem [{\citenamefont {Sorscher}\ \emph {et~al.}(2022)\citenamefont
  {Sorscher}, \citenamefont {Ganguli},\ and\ \citenamefont
  {Sompolinsky}}]{Sorscher2022-da}%
  \BibitemOpen
  \bibfield  {author} {\bibinfo {author} {\bibfnamefont {B.}~\bibnamefont
  {Sorscher}}, \bibinfo {author} {\bibfnamefont {S.}~\bibnamefont {Ganguli}},\
  and\ \bibinfo {author} {\bibfnamefont {H.}~\bibnamefont {Sompolinsky}},\
  }\bibfield  {title} {\bibinfo {title} {Neural representational geometry
  underlies few-shot concept learning},\ }\href
  {https://doi.org/10.1073/pnas.2200800119} {\bibfield  {journal} {\bibinfo
  {journal} {Proc. Natl Acad. Sci. USA}\ }\textbf {\bibinfo {volume} {119}},\
  \bibinfo {pages} {e2200800119} (\bibinfo {year} {2022})}\BibitemShut
  {NoStop}%
\bibitem [{\citenamefont {Rust}\ and\ \citenamefont
  {Cohen}(2022)}]{Rust2022-av}%
  \BibitemOpen
  \bibfield  {author} {\bibinfo {author} {\bibfnamefont {N.~C.}\ \bibnamefont
  {Rust}}\ and\ \bibinfo {author} {\bibfnamefont {M.~R.}\ \bibnamefont
  {Cohen}},\ }\bibfield  {title} {\bibinfo {title} {Priority coding in the
  visual system},\ }\href {https://doi.org/10.1038/s41583-022-00582-9}
  {\bibfield  {journal} {\bibinfo  {journal} {Nat. Rev. Neurosci.}\ }\textbf
  {\bibinfo {volume} {23}},\ \bibinfo {pages} {376} (\bibinfo {year}
  {2022})}\BibitemShut {NoStop}%
\bibitem [{\citenamefont {Nogueira}\ \emph {et~al.}(2023)\citenamefont
  {Nogueira}, \citenamefont {Rodgers}, \citenamefont {Bruno},\ and\
  \citenamefont {Fusi}}]{Nogueira2023-se}%
  \BibitemOpen
  \bibfield  {author} {\bibinfo {author} {\bibfnamefont {R.}~\bibnamefont
  {Nogueira}}, \bibinfo {author} {\bibfnamefont {C.~C.}\ \bibnamefont
  {Rodgers}}, \bibinfo {author} {\bibfnamefont {R.~M.}\ \bibnamefont {Bruno}},\
  and\ \bibinfo {author} {\bibfnamefont {S.}~\bibnamefont {Fusi}},\ }\bibfield
  {title} {\bibinfo {title} {The geometry of cortical representations of touch
  in rodents},\ }\href {https://doi.org/10.1038/s41593-022-01237-9} {\bibfield
  {journal} {\bibinfo  {journal} {Nat. Neurosci.}\ }\textbf {\bibinfo {volume}
  {26}},\ \bibinfo {pages} {239} (\bibinfo {year} {2023})}\BibitemShut
  {NoStop}%
\bibitem [{\citenamefont {Schneider}\ \emph {et~al.}(2023)\citenamefont
  {Schneider}, \citenamefont {Lee},\ and\ \citenamefont
  {Mathis}}]{Schneider2023-cj}%
  \BibitemOpen
  \bibfield  {author} {\bibinfo {author} {\bibfnamefont {S.}~\bibnamefont
  {Schneider}}, \bibinfo {author} {\bibfnamefont {J.~H.}\ \bibnamefont {Lee}},\
  and\ \bibinfo {author} {\bibfnamefont {M.~W.}\ \bibnamefont {Mathis}},\
  }\bibfield  {title} {\bibinfo {title} {Learnable latent embeddings for joint
  behavioural and neural analysis},\ }\href
  {https://doi.org/10.1038/s41586-023-06031-6} {\bibfield  {journal} {\bibinfo
  {journal} {Nature}\ }\textbf {\bibinfo {volume} {617}},\ \bibinfo {pages}
  {360} (\bibinfo {year} {2023})}\BibitemShut {NoStop}%
\bibitem [{\citenamefont {Manley}\ \emph {et~al.}(2024)\citenamefont {Manley},
  \citenamefont {Lu}, \citenamefont {Barber}, \citenamefont {Demas},
  \citenamefont {Kim}, \citenamefont {Meyer}, \citenamefont {Traub},\ and\
  \citenamefont {Vaziri}}]{Manley2024-bn}%
  \BibitemOpen
  \bibfield  {author} {\bibinfo {author} {\bibfnamefont {J.}~\bibnamefont
  {Manley}}, \bibinfo {author} {\bibfnamefont {S.}~\bibnamefont {Lu}}, \bibinfo
  {author} {\bibfnamefont {K.}~\bibnamefont {Barber}}, \bibinfo {author}
  {\bibfnamefont {J.}~\bibnamefont {Demas}}, \bibinfo {author} {\bibfnamefont
  {H.}~\bibnamefont {Kim}}, \bibinfo {author} {\bibfnamefont {D.}~\bibnamefont
  {Meyer}}, \bibinfo {author} {\bibfnamefont {F.~M.}\ \bibnamefont {Traub}},\
  and\ \bibinfo {author} {\bibfnamefont {A.}~\bibnamefont {Vaziri}},\
  }\bibfield  {title} {\bibinfo {title} {Simultaneous, cortex-wide dynamics of
  up to 1 million neurons reveal unbounded scaling of dimensionality with
  neuron number},\ }\href {https://doi.org/10.1016/j.neuron.2024.02.011}
  {\bibfield  {journal} {\bibinfo  {journal} {Neuron}\ }\textbf {\bibinfo
  {volume} {112}},\ \bibinfo {pages} {1694} (\bibinfo {year}
  {2024})}\BibitemShut {NoStop}%
\bibitem [{\citenamefont {Poole}\ \emph {et~al.}(2016)\citenamefont {Poole},
  \citenamefont {Lahiri}, \citenamefont {Raghu}, \citenamefont
  {Sohl-Dickstein},\ and\ \citenamefont {Ganguli}}]{Poole2016-wo}%
  \BibitemOpen
  \bibfield  {author} {\bibinfo {author} {\bibfnamefont {B.}~\bibnamefont
  {Poole}}, \bibinfo {author} {\bibfnamefont {S.}~\bibnamefont {Lahiri}},
  \bibinfo {author} {\bibfnamefont {M.}~\bibnamefont {Raghu}}, \bibinfo
  {author} {\bibfnamefont {J.~N.}\ \bibnamefont {Sohl-Dickstein}},\ and\
  \bibinfo {author} {\bibfnamefont {S.}~\bibnamefont {Ganguli}},\ }\bibfield
  {title} {\bibinfo {title} {Exponential expressivity in deep neural networks
  through transient chaos},\ }\href@noop {} {\bibfield  {journal} {\bibinfo
  {journal} {Adv. Neural Inf. Proc. Sys.}\ }\textbf {\bibinfo {volume} {29}},\
  \bibinfo {pages} {3368} (\bibinfo {year} {2016})}\BibitemShut {NoStop}%
\bibitem [{\citenamefont {Schoenholz}\ \emph {et~al.}(2017)\citenamefont
  {Schoenholz}, \citenamefont {Gilmer}, \citenamefont {Ganguli},\ and\
  \citenamefont {Sohl-Dickstein}}]{Schoenholz2016-ll}%
  \BibitemOpen
  \bibfield  {author} {\bibinfo {author} {\bibfnamefont {S.~S.}\ \bibnamefont
  {Schoenholz}}, \bibinfo {author} {\bibfnamefont {J.}~\bibnamefont {Gilmer}},
  \bibinfo {author} {\bibfnamefont {S.}~\bibnamefont {Ganguli}},\ and\ \bibinfo
  {author} {\bibfnamefont {J.}~\bibnamefont {Sohl-Dickstein}},\ }\bibfield
  {title} {\bibinfo {title} {Deep information propagation},\ }in\ \href@noop {}
  {\emph {\bibinfo {booktitle} {International Conference on Learning
  Representations}}},\ \bibinfo {editor} {edited by\ \bibinfo {editor}
  {\bibnamefont {""}}}\ (\bibinfo {year} {2017})\BibitemShut {NoStop}%
\bibitem [{\citenamefont {Pennington}\ \emph {et~al.}(2017)\citenamefont
  {Pennington}, \citenamefont {Schoenholz},\ and\ \citenamefont
  {Ganguli}}]{Pennington2017-hk}%
  \BibitemOpen
  \bibfield  {author} {\bibinfo {author} {\bibfnamefont {J.}~\bibnamefont
  {Pennington}}, \bibinfo {author} {\bibfnamefont {S.}~\bibnamefont
  {Schoenholz}},\ and\ \bibinfo {author} {\bibfnamefont {S.}~\bibnamefont
  {Ganguli}},\ }\bibfield  {title} {\bibinfo {title} {Resurrecting the sigmoid
  in deep learning through dynamical isometry: theory and practice},\
  }\href@noop {} {\bibfield  {journal} {\bibinfo  {journal} {Adv. Neural Inf.
  Proc. Sys.}\ }\textbf {\bibinfo {volume} {30}},\ \bibinfo {pages} {4788}
  (\bibinfo {year} {2017})}\BibitemShut {NoStop}%
\bibitem [{\citenamefont {Kaplan}\ \emph {et~al.}(2020)\citenamefont {Kaplan},
  \citenamefont {McCandlish}, \citenamefont {Henighan}, \citenamefont {Brown},
  \citenamefont {Chess}, \citenamefont {Child}, \citenamefont {Gray},
  \citenamefont {Radford}, \citenamefont {Wu},\ and\ \citenamefont
  {Amodei}}]{Kaplan2020-zt}%
  \BibitemOpen
  \bibfield  {author} {\bibinfo {author} {\bibfnamefont {J.}~\bibnamefont
  {Kaplan}}, \bibinfo {author} {\bibfnamefont {S.}~\bibnamefont {McCandlish}},
  \bibinfo {author} {\bibfnamefont {T.}~\bibnamefont {Henighan}}, \bibinfo
  {author} {\bibfnamefont {T.~B.}\ \bibnamefont {Brown}}, \bibinfo {author}
  {\bibfnamefont {B.}~\bibnamefont {Chess}}, \bibinfo {author} {\bibfnamefont
  {R.}~\bibnamefont {Child}}, \bibinfo {author} {\bibfnamefont
  {S.}~\bibnamefont {Gray}}, \bibinfo {author} {\bibfnamefont {A.}~\bibnamefont
  {Radford}}, \bibinfo {author} {\bibfnamefont {J.}~\bibnamefont {Wu}},\ and\
  \bibinfo {author} {\bibfnamefont {D.}~\bibnamefont {Amodei}},\ }\href@noop {}
  {\bibinfo {title} {Scaling laws for neural language models}},\ \bibinfo
  {howpublished} {Preprint at \url{https://arxiv.org/abs/2001.08361}} (\bibinfo
  {year} {2020})\BibitemShut {NoStop}%
\bibitem [{\citenamefont {Nassar}\ \emph {et~al.}(2020)\citenamefont {Nassar},
  \citenamefont {Sokól}, \citenamefont {Chung}, \citenamefont {Harris},\ and\
  \citenamefont {Park}}]{Nassar2020-pz}%
  \BibitemOpen
  \bibfield  {author} {\bibinfo {author} {\bibfnamefont {J.}~\bibnamefont
  {Nassar}}, \bibinfo {author} {\bibfnamefont {P.~A.}\ \bibnamefont {Sokól}},
  \bibinfo {author} {\bibfnamefont {S.}~\bibnamefont {Chung}}, \bibinfo
  {author} {\bibfnamefont {K.}~\bibnamefont {Harris}},\ and\ \bibinfo {author}
  {\bibfnamefont {I.~M.}\ \bibnamefont {Park}},\ }\bibfield  {title} {\bibinfo
  {title} {On 1/n neural representation and robustness},\ }\href@noop {}
  {\bibfield  {journal} {\bibinfo  {journal} {Adv. Neural Inf. Proc. Sys.}\
  }\textbf {\bibinfo {volume} {33}},\ \bibinfo {pages} {6211} (\bibinfo {year}
  {2020})}\BibitemShut {NoStop}%
\bibitem [{\citenamefont {Chen}\ \emph {et~al.}(2022)\citenamefont {Chen},
  \citenamefont {Scherr},\ and\ \citenamefont {Maass}}]{Chen2022-ts}%
  \BibitemOpen
  \bibfield  {author} {\bibinfo {author} {\bibfnamefont {G.}~\bibnamefont
  {Chen}}, \bibinfo {author} {\bibfnamefont {F.}~\bibnamefont {Scherr}},\ and\
  \bibinfo {author} {\bibfnamefont {W.}~\bibnamefont {Maass}},\ }\bibfield
  {title} {\bibinfo {title} {A data-based large-scale model for primary visual
  cortex enables brain-like robust and versatile visual processing},\ }\href
  {https://doi.org/10.1126/sciadv.abq7592} {\bibfield  {journal} {\bibinfo
  {journal} {Sci. Adv.}\ }\textbf {\bibinfo {volume} {8}},\ \bibinfo {pages}
  {eabq7592} (\bibinfo {year} {2022})}\BibitemShut {NoStop}%
\bibitem [{\citenamefont {Johnston}\ and\ \citenamefont
  {Fusi}(2023)}]{Johnston2023-ov}%
  \BibitemOpen
  \bibfield  {author} {\bibinfo {author} {\bibfnamefont {W.~J.}\ \bibnamefont
  {Johnston}}\ and\ \bibinfo {author} {\bibfnamefont {S.}~\bibnamefont
  {Fusi}},\ }\bibfield  {title} {\bibinfo {title} {Abstract representations
  emerge naturally in neural networks trained to perform multiple tasks},\
  }\href {https://doi.org/10.1038/s41467-023-36583-0} {\bibfield  {journal}
  {\bibinfo  {journal} {Nat. Commun.}\ }\textbf {\bibinfo {volume} {14}},\
  \bibinfo {pages} {1040} (\bibinfo {year} {2023})}\BibitemShut {NoStop}%
\bibitem [{\citenamefont {Joseph}\ \emph {et~al.}(2023)\citenamefont {Joseph},
  \citenamefont {Agrawal}, \citenamefont {A.},\ and\ \citenamefont
  {Richards}}]{Joseph2023-ct}%
  \BibitemOpen
  \bibfield  {author} {\bibinfo {author} {\bibfnamefont {S.}~\bibnamefont
  {Joseph}}, \bibinfo {author} {\bibfnamefont {K.~K.}\ \bibnamefont {Agrawal}},
  \bibinfo {author} {\bibfnamefont {G.}~\bibnamefont {A.}},\ and\ \bibinfo
  {author} {\bibfnamefont {B.~A.}\ \bibnamefont {Richards}},\ }\bibfield
  {title} {\bibinfo {title} {On the information geometry of vision
  transformers},\ }in\ \href@noop {} {\emph {\bibinfo {booktitle} {NeurIPS 2023
  Workshop on Symmetry and Geometry in Neural Representations}}},\ \bibinfo
  {editor} {edited by\ \bibinfo {editor} {\bibnamefont {""}}}\ (\bibinfo {year}
  {2023})\BibitemShut {NoStop}%
\bibitem [{\citenamefont {Beggs}\ and\ \citenamefont
  {Plenz}(2003)}]{Beggs2003-ac}%
  \BibitemOpen
  \bibfield  {author} {\bibinfo {author} {\bibfnamefont {J.~M.}\ \bibnamefont
  {Beggs}}\ and\ \bibinfo {author} {\bibfnamefont {D.}~\bibnamefont {Plenz}},\
  }\bibfield  {title} {\bibinfo {title} {Neuronal avalanches in neocortical
  circuits},\ }\href {https://doi.org/10.1523/JNEUROSCI.23-35-11167.2003}
  {\bibfield  {journal} {\bibinfo  {journal} {J. Neurosci.}\ }\textbf {\bibinfo
  {volume} {23}},\ \bibinfo {pages} {11167} (\bibinfo {year}
  {2003})}\BibitemShut {NoStop}%
\bibitem [{\citenamefont {de~Arcangelis}\ \emph {et~al.}(2006)\citenamefont
  {de~Arcangelis}, \citenamefont {Perrone-Capano},\ and\ \citenamefont
  {Herrmann}}]{De_Arcangelis2006-wx}%
  \BibitemOpen
  \bibfield  {author} {\bibinfo {author} {\bibfnamefont {L.}~\bibnamefont
  {de~Arcangelis}}, \bibinfo {author} {\bibfnamefont {C.}~\bibnamefont
  {Perrone-Capano}},\ and\ \bibinfo {author} {\bibfnamefont {H.~J.}\
  \bibnamefont {Herrmann}},\ }\bibfield  {title} {\bibinfo {title}
  {Self-organized criticality model for brain plasticity},\ }\href
  {https://doi.org/10.1103/PhysRevLett.96.028107} {\bibfield  {journal}
  {\bibinfo  {journal} {Phys. Rev. Lett.}\ }\textbf {\bibinfo {volume} {96}},\
  \bibinfo {pages} {028107} (\bibinfo {year} {2006})}\BibitemShut {NoStop}%
\bibitem [{\citenamefont {Tkačik}\ \emph {et~al.}(2015)\citenamefont
  {Tkačik}, \citenamefont {Mora}, \citenamefont {Marre}, \citenamefont
  {Amodei}, \citenamefont {Palmer}, \citenamefont {Berry},\ and\ \citenamefont
  {Bialek}}]{Tkacik2015-ly}%
  \BibitemOpen
  \bibfield  {author} {\bibinfo {author} {\bibfnamefont {G.}~\bibnamefont
  {Tkačik}}, \bibinfo {author} {\bibfnamefont {T.}~\bibnamefont {Mora}},
  \bibinfo {author} {\bibfnamefont {O.}~\bibnamefont {Marre}}, \bibinfo
  {author} {\bibfnamefont {D.}~\bibnamefont {Amodei}}, \bibinfo {author}
  {\bibfnamefont {S.~E.}\ \bibnamefont {Palmer}}, \bibinfo {author}
  {\bibfnamefont {M.~J.}\ \bibnamefont {Berry}, \bibfnamefont {2nd}},\ and\
  \bibinfo {author} {\bibfnamefont {W.}~\bibnamefont {Bialek}},\ }\bibfield
  {title} {\bibinfo {title} {Thermodynamics and signatures of criticality in a
  network of neurons},\ }\href {https://doi.org/10.1073/pnas.1514188112}
  {\bibfield  {journal} {\bibinfo  {journal} {Proc. Natl Acad. Sci. USA}\
  }\textbf {\bibinfo {volume} {112}},\ \bibinfo {pages} {11508} (\bibinfo
  {year} {2015})}\BibitemShut {NoStop}%
\bibitem [{\citenamefont {Dahmen}\ \emph {et~al.}(2019)\citenamefont {Dahmen},
  \citenamefont {Grün}, \citenamefont {Diesmann},\ and\ \citenamefont
  {Helias}}]{Dahmen2019-qr}%
  \BibitemOpen
  \bibfield  {author} {\bibinfo {author} {\bibfnamefont {D.}~\bibnamefont
  {Dahmen}}, \bibinfo {author} {\bibfnamefont {S.}~\bibnamefont {Grün}},
  \bibinfo {author} {\bibfnamefont {M.}~\bibnamefont {Diesmann}},\ and\
  \bibinfo {author} {\bibfnamefont {M.}~\bibnamefont {Helias}},\ }\bibfield
  {title} {\bibinfo {title} {Second type of criticality in the brain uncovers
  rich multiple-neuron dynamics},\ }\href
  {https://doi.org/10.1073/pnas.1818972116} {\bibfield  {journal} {\bibinfo
  {journal} {Proc. Natl Acad. Sci. USA}\ }\textbf {\bibinfo {volume} {116}},\
  \bibinfo {pages} {13051} (\bibinfo {year} {2019})}\BibitemShut {NoStop}%
\bibitem [{\citenamefont {Fosque}\ \emph {et~al.}(2021)\citenamefont {Fosque},
  \citenamefont {Williams-García}, \citenamefont {Beggs},\ and\ \citenamefont
  {Ortiz}}]{Fosque2021-fw}%
  \BibitemOpen
  \bibfield  {author} {\bibinfo {author} {\bibfnamefont {L.~J.}\ \bibnamefont
  {Fosque}}, \bibinfo {author} {\bibfnamefont {R.~V.}\ \bibnamefont
  {Williams-García}}, \bibinfo {author} {\bibfnamefont {J.~M.}\ \bibnamefont
  {Beggs}},\ and\ \bibinfo {author} {\bibfnamefont {G.}~\bibnamefont {Ortiz}},\
  }\bibfield  {title} {\bibinfo {title} {Evidence for quasicritical brain
  dynamics},\ }\href {https://doi.org/10.1103/PhysRevLett.126.098101}
  {\bibfield  {journal} {\bibinfo  {journal} {Phys. Rev. Lett.}\ }\textbf
  {\bibinfo {volume} {126}},\ \bibinfo {pages} {098101} (\bibinfo {year}
  {2021})}\BibitemShut {NoStop}%
\bibitem [{\citenamefont {O'Byrne}\ and\ \citenamefont
  {Jerbi}(2022)}]{OByrne2022-no}%
  \BibitemOpen
  \bibfield  {author} {\bibinfo {author} {\bibfnamefont {J.}~\bibnamefont
  {O'Byrne}}\ and\ \bibinfo {author} {\bibfnamefont {K.}~\bibnamefont
  {Jerbi}},\ }\bibfield  {title} {\bibinfo {title} {How critical is brain
  criticality?},\ }\href {https://doi.org/10.1016/j.tins.2022.08.007}
  {\bibfield  {journal} {\bibinfo  {journal} {Trends Neurosci.}\ }\textbf
  {\bibinfo {volume} {45}},\ \bibinfo {pages} {820} (\bibinfo {year}
  {2022})}\BibitemShut {NoStop}%
\bibitem [{\citenamefont {Morales}\ \emph {et~al.}(2023)\citenamefont
  {Morales}, \citenamefont {di~Santo},\ and\ \citenamefont
  {Muñoz}}]{Morales2023-ii}%
  \BibitemOpen
  \bibfield  {author} {\bibinfo {author} {\bibfnamefont {G.~B.}\ \bibnamefont
  {Morales}}, \bibinfo {author} {\bibfnamefont {S.}~\bibnamefont {di~Santo}},\
  and\ \bibinfo {author} {\bibfnamefont {M.~A.}\ \bibnamefont {Muñoz}},\
  }\bibfield  {title} {\bibinfo {title} {Quasiuniversal scaling in mouse-brain
  neuronal activity stems from edge-of-instability critical dynamics},\ }\href
  {https://doi.org/10.1073/pnas.2208998120} {\bibfield  {journal} {\bibinfo
  {journal} {Proc. Natl Acad. Sci. USA}\ }\textbf {\bibinfo {volume} {120}},\
  \bibinfo {pages} {e2208998120} (\bibinfo {year} {2023})}\BibitemShut
  {NoStop}%
\bibitem [{\citenamefont {Seung}\ and\ \citenamefont
  {Sompolinsky}(1993)}]{Seung1993-qt}%
  \BibitemOpen
  \bibfield  {author} {\bibinfo {author} {\bibfnamefont {H.~S.}\ \bibnamefont
  {Seung}}\ and\ \bibinfo {author} {\bibfnamefont {H.}~\bibnamefont
  {Sompolinsky}},\ }\bibfield  {title} {\bibinfo {title} {Simple models for
  reading neuronal population codes},\ }\href
  {https://doi.org/10.1073/pnas.90.22.10749} {\bibfield  {journal} {\bibinfo
  {journal} {Proc. Natl Acad. Sci. USA}\ }\textbf {\bibinfo {volume} {90}},\
  \bibinfo {pages} {10749} (\bibinfo {year} {1993})}\BibitemShut {NoStop}%
\bibitem [{\citenamefont {Brunel}\ and\ \citenamefont
  {Nadal}(1998)}]{Brunel1998-wb}%
  \BibitemOpen
  \bibfield  {author} {\bibinfo {author} {\bibfnamefont {N.}~\bibnamefont
  {Brunel}}\ and\ \bibinfo {author} {\bibfnamefont {J.~P.}\ \bibnamefont
  {Nadal}},\ }\bibfield  {title} {\bibinfo {title} {Mutual information, fisher
  information, and population coding},\ }\href
  {https://doi.org/10.1162/089976698300017115} {\bibfield  {journal} {\bibinfo
  {journal} {Neural Comput.}\ }\textbf {\bibinfo {volume} {10}},\ \bibinfo
  {pages} {1731} (\bibinfo {year} {1998})}\BibitemShut {NoStop}%
\bibitem [{\citenamefont {Deneve}\ \emph {et~al.}(1999)\citenamefont {Deneve},
  \citenamefont {Latham},\ and\ \citenamefont {Pouget}}]{Deneve1999-sf}%
  \BibitemOpen
  \bibfield  {author} {\bibinfo {author} {\bibfnamefont {S.}~\bibnamefont
  {Deneve}}, \bibinfo {author} {\bibfnamefont {P.~E.}\ \bibnamefont {Latham}},\
  and\ \bibinfo {author} {\bibfnamefont {A.}~\bibnamefont {Pouget}},\
  }\bibfield  {title} {\bibinfo {title} {Reading population codes: a neural
  implementation of ideal observers},\ }\href {https://doi.org/10.1038/11205}
  {\bibfield  {journal} {\bibinfo  {journal} {Nat. Neurosci.}\ }\textbf
  {\bibinfo {volume} {2}},\ \bibinfo {pages} {740} (\bibinfo {year}
  {1999})}\BibitemShut {NoStop}%
\bibitem [{\citenamefont {Sompolinsky}\ \emph {et~al.}(2001)\citenamefont
  {Sompolinsky}, \citenamefont {Yoon}, \citenamefont {Kang},\ and\
  \citenamefont {Shamir}}]{Sompolinsky2001-dl}%
  \BibitemOpen
  \bibfield  {author} {\bibinfo {author} {\bibfnamefont {H.}~\bibnamefont
  {Sompolinsky}}, \bibinfo {author} {\bibfnamefont {H.}~\bibnamefont {Yoon}},
  \bibinfo {author} {\bibfnamefont {K.}~\bibnamefont {Kang}},\ and\ \bibinfo
  {author} {\bibfnamefont {M.}~\bibnamefont {Shamir}},\ }\bibfield  {title}
  {\bibinfo {title} {Population coding in neuronal systems with correlated
  noise},\ }\href {https://doi.org/10.1103/PhysRevE.64.051904} {\bibfield
  {journal} {\bibinfo  {journal} {Phys. Rev. E Stat. Nonlin. Soft Matter
  Phys.}\ }\textbf {\bibinfo {volume} {64}},\ \bibinfo {pages} {051904}
  (\bibinfo {year} {2001})}\BibitemShut {NoStop}%
\bibitem [{\citenamefont {Quian~Quiroga}\ and\ \citenamefont
  {Panzeri}(2009)}]{Quian_Quiroga2009-rz}%
  \BibitemOpen
  \bibfield  {author} {\bibinfo {author} {\bibfnamefont {R.}~\bibnamefont
  {Quian~Quiroga}}\ and\ \bibinfo {author} {\bibfnamefont {S.}~\bibnamefont
  {Panzeri}},\ }\bibfield  {title} {\bibinfo {title} {Extracting information
  from neuronal populations: information theory and decoding approaches},\
  }\href {https://doi.org/10.1038/nrn2578} {\bibfield  {journal} {\bibinfo
  {journal} {Nat. Rev. Neurosci.}\ }\textbf {\bibinfo {volume} {10}},\ \bibinfo
  {pages} {173} (\bibinfo {year} {2009})}\BibitemShut {NoStop}%
\bibitem [{\citenamefont {Moreno-Bote}\ \emph {et~al.}(2014)\citenamefont
  {Moreno-Bote}, \citenamefont {Beck}, \citenamefont {Kanitscheider},
  \citenamefont {Pitkow}, \citenamefont {Latham},\ and\ \citenamefont
  {Pouget}}]{Moreno-Bote2014-dh}%
  \BibitemOpen
  \bibfield  {author} {\bibinfo {author} {\bibfnamefont {R.}~\bibnamefont
  {Moreno-Bote}}, \bibinfo {author} {\bibfnamefont {J.}~\bibnamefont {Beck}},
  \bibinfo {author} {\bibfnamefont {I.}~\bibnamefont {Kanitscheider}}, \bibinfo
  {author} {\bibfnamefont {X.}~\bibnamefont {Pitkow}}, \bibinfo {author}
  {\bibfnamefont {P.}~\bibnamefont {Latham}},\ and\ \bibinfo {author}
  {\bibfnamefont {A.}~\bibnamefont {Pouget}},\ }\bibfield  {title} {\bibinfo
  {title} {Information-limiting correlations},\ }\href
  {https://doi.org/10.1038/nn.3807} {\bibfield  {journal} {\bibinfo  {journal}
  {Nat. Neurosci.}\ }\textbf {\bibinfo {volume} {17}},\ \bibinfo {pages} {1410}
  (\bibinfo {year} {2014})}\BibitemShut {NoStop}%
\bibitem [{\citenamefont {Cover}(1999)}]{cover1999elements}%
  \BibitemOpen
  \bibfield  {author} {\bibinfo {author} {\bibfnamefont {T.~M.}\ \bibnamefont
  {Cover}},\ }\href@noop {} {\emph {\bibinfo {title} {Elements of information
  theory}}}\ (\bibinfo  {publisher} {John Wiley \& Sons},\ \bibinfo {year}
  {1999})\BibitemShut {NoStop}%
\bibitem [{\citenamefont {Kubo}(1966)}]{Kubo1966-rg}%
  \BibitemOpen
  \bibfield  {author} {\bibinfo {author} {\bibfnamefont {R.}~\bibnamefont
  {Kubo}},\ }\bibfield  {title} {\bibinfo {title} {The fluctuation-dissipation
  theorem},\ }\href {https://doi.org/10.1088/0034-4885/29/1/306} {\bibfield
  {journal} {\bibinfo  {journal} {Rep. Prog. Phys.}\ }\textbf {\bibinfo
  {volume} {29}},\ \bibinfo {pages} {255} (\bibinfo {year} {1966})}\BibitemShut
  {NoStop}%
\bibitem [{\citenamefont {Laughlin}\ \emph {et~al.}(1998)\citenamefont
  {Laughlin}, \citenamefont {de~Ruyter~van Steveninck},\ and\ \citenamefont
  {Anderson}}]{Laughlin1998-yu}%
  \BibitemOpen
  \bibfield  {author} {\bibinfo {author} {\bibfnamefont {S.~B.}\ \bibnamefont
  {Laughlin}}, \bibinfo {author} {\bibfnamefont {R.~R.}\ \bibnamefont
  {de~Ruyter~van Steveninck}},\ and\ \bibinfo {author} {\bibfnamefont {J.~C.}\
  \bibnamefont {Anderson}},\ }\bibfield  {title} {\bibinfo {title} {The
  metabolic cost of neural information},\ }\href {https://doi.org/10.1038/236}
  {\bibfield  {journal} {\bibinfo  {journal} {Nat. Neurosci.}\ }\textbf
  {\bibinfo {volume} {1}},\ \bibinfo {pages} {36} (\bibinfo {year}
  {1998})}\BibitemShut {NoStop}%
\bibitem [{\citenamefont {Lennie}(2003)}]{Lennie2003-td}%
  \BibitemOpen
  \bibfield  {author} {\bibinfo {author} {\bibfnamefont {P.}~\bibnamefont
  {Lennie}},\ }\bibfield  {title} {\bibinfo {title} {The cost of cortical
  computation},\ }\href {https://doi.org/10.1016/s0960-9822(03)00135-0}
  {\bibfield  {journal} {\bibinfo  {journal} {Curr. Biol.}\ }\textbf {\bibinfo
  {volume} {13}},\ \bibinfo {pages} {493} (\bibinfo {year} {2003})}\BibitemShut
  {NoStop}%
\bibitem [{\citenamefont {Kong}\ \emph {et~al.}(2022)\citenamefont {Kong},
  \citenamefont {Margalit}, \citenamefont {Gardner},\ and\ \citenamefont
  {Norcia}}]{Kong2022-fh}%
  \BibitemOpen
  \bibfield  {author} {\bibinfo {author} {\bibfnamefont {N.~C.~L.}\
  \bibnamefont {Kong}}, \bibinfo {author} {\bibfnamefont {E.}~\bibnamefont
  {Margalit}}, \bibinfo {author} {\bibfnamefont {J.~L.}\ \bibnamefont
  {Gardner}},\ and\ \bibinfo {author} {\bibfnamefont {A.~M.}\ \bibnamefont
  {Norcia}},\ }\bibfield  {title} {\bibinfo {title} {Increasing neural network
  robustness improves match to macaque {V1} eigenspectrum, spatial frequency
  preference and predictivity},\ }\href
  {https://doi.org/10.1371/journal.pcbi.1009739} {\bibfield  {journal}
  {\bibinfo  {journal} {PLoS Comput. Biol.}\ }\textbf {\bibinfo {volume}
  {18}},\ \bibinfo {pages} {e1009739} (\bibinfo {year} {2022})}\BibitemShut
  {NoStop}%
\bibitem [{\citenamefont {Hu}\ and\ \citenamefont
  {Sompolinsky}(2022)}]{Hu2022-ib}%
  \BibitemOpen
  \bibfield  {author} {\bibinfo {author} {\bibfnamefont {Y.}~\bibnamefont
  {Hu}}\ and\ \bibinfo {author} {\bibfnamefont {H.}~\bibnamefont
  {Sompolinsky}},\ }\bibfield  {title} {\bibinfo {title} {The spectrum of
  covariance matrices of randomly connected recurrent neuronal networks with
  linear dynamics},\ }\href {https://doi.org/10.1371/journal.pcbi.1010327}
  {\bibfield  {journal} {\bibinfo  {journal} {PLoS Comput. Biol.}\ }\textbf
  {\bibinfo {volume} {18}},\ \bibinfo {pages} {e1010327} (\bibinfo {year}
  {2022})}\BibitemShut {NoStop}%
\bibitem [{\citenamefont {Huang}\ \emph {et~al.}(2022)\citenamefont {Huang},
  \citenamefont {Pouget},\ and\ \citenamefont {Doiron}}]{Huang2022-sa}%
  \BibitemOpen
  \bibfield  {author} {\bibinfo {author} {\bibfnamefont {C.}~\bibnamefont
  {Huang}}, \bibinfo {author} {\bibfnamefont {A.}~\bibnamefont {Pouget}},\ and\
  \bibinfo {author} {\bibfnamefont {B.}~\bibnamefont {Doiron}},\ }\bibfield
  {title} {\bibinfo {title} {Internally generated population activity in
  cortical networks hinders information transmission},\ }\href
  {https://doi.org/10.1126/sciadv.abg5244} {\bibfield  {journal} {\bibinfo
  {journal} {Sci. Adv.}\ }\textbf {\bibinfo {volume} {8}},\ \bibinfo {pages}
  {eabg5244} (\bibinfo {year} {2022})}\BibitemShut {NoStop}%
\bibitem [{\citenamefont {Wardak}\ and\ \citenamefont
  {Gong}(2022)}]{Wardak2022-xu}%
  \BibitemOpen
  \bibfield  {author} {\bibinfo {author} {\bibfnamefont {A.}~\bibnamefont
  {Wardak}}\ and\ \bibinfo {author} {\bibfnamefont {P.}~\bibnamefont {Gong}},\
  }\bibfield  {title} {\bibinfo {title} {Extended anderson criticality in
  heavy-tailed neural networks},\ }\href
  {https://doi.org/10.1103/PhysRevLett.129.048103} {\bibfield  {journal}
  {\bibinfo  {journal} {Phys. Rev. Lett.}\ }\textbf {\bibinfo {volume} {129}},\
  \bibinfo {pages} {048103} (\bibinfo {year} {2022})}\BibitemShut {NoStop}%
\bibitem [{\citenamefont {Vyas}\ \emph {et~al.}(2020)\citenamefont {Vyas},
  \citenamefont {Golub}, \citenamefont {Sussillo},\ and\ \citenamefont
  {Shenoy}}]{Vyas2020-eu}%
  \BibitemOpen
  \bibfield  {author} {\bibinfo {author} {\bibfnamefont {S.}~\bibnamefont
  {Vyas}}, \bibinfo {author} {\bibfnamefont {M.~D.}\ \bibnamefont {Golub}},
  \bibinfo {author} {\bibfnamefont {D.}~\bibnamefont {Sussillo}},\ and\
  \bibinfo {author} {\bibfnamefont {K.~V.}\ \bibnamefont {Shenoy}},\ }\bibfield
   {title} {\bibinfo {title} {Computation through neural population dynamics},\
  }\href {https://doi.org/10.1146/annurev-neuro-092619-094115} {\bibfield
  {journal} {\bibinfo  {journal} {Annu. Rev. Neurosci.}\ }\textbf {\bibinfo
  {volume} {43}},\ \bibinfo {pages} {249} (\bibinfo {year} {2020})}\BibitemShut
  {NoStop}%
\bibitem [{\citenamefont {Dubreuil}\ \emph {et~al.}(2022)\citenamefont
  {Dubreuil}, \citenamefont {Valente}, \citenamefont {Beiran}, \citenamefont
  {Mastrogiuseppe},\ and\ \citenamefont {Ostojic}}]{Dubreuil2022-qt}%
  \BibitemOpen
  \bibfield  {author} {\bibinfo {author} {\bibfnamefont {A.}~\bibnamefont
  {Dubreuil}}, \bibinfo {author} {\bibfnamefont {A.}~\bibnamefont {Valente}},
  \bibinfo {author} {\bibfnamefont {M.}~\bibnamefont {Beiran}}, \bibinfo
  {author} {\bibfnamefont {F.}~\bibnamefont {Mastrogiuseppe}},\ and\ \bibinfo
  {author} {\bibfnamefont {S.}~\bibnamefont {Ostojic}},\ }\bibfield  {title}
  {\bibinfo {title} {The role of population structure in computations through
  neural dynamics},\ }\href {https://doi.org/10.1038/s41593-022-01088-4}
  {\bibfield  {journal} {\bibinfo  {journal} {Nat. Neurosci.}\ }\textbf
  {\bibinfo {volume} {25}},\ \bibinfo {pages} {783} (\bibinfo {year}
  {2022})}\BibitemShut {NoStop}%
\bibitem [{\citenamefont {De}\ and\ \citenamefont
  {Chaudhuri}(2023)}]{De2023-su}%
  \BibitemOpen
  \bibfield  {author} {\bibinfo {author} {\bibfnamefont {A.}~\bibnamefont
  {De}}\ and\ \bibinfo {author} {\bibfnamefont {R.}~\bibnamefont {Chaudhuri}},\
  }\bibfield  {title} {\bibinfo {title} {Common population codes produce
  extremely nonlinear neural manifolds},\ }\href
  {https://doi.org/10.1073/pnas.2305853120} {\bibfield  {journal} {\bibinfo
  {journal} {Proc. Natl Acad. Sci. USA}\ }\textbf {\bibinfo {volume} {120}},\
  \bibinfo {pages} {e2305853120} (\bibinfo {year} {2023})}\BibitemShut
  {NoStop}%
\bibitem [{\citenamefont {Amari}\ and\ \citenamefont
  {Nagaoka}(2000)}]{amari2000methods}%
  \BibitemOpen
  \bibfield  {author} {\bibinfo {author} {\bibfnamefont {S.-i.}\ \bibnamefont
  {Amari}}\ and\ \bibinfo {author} {\bibfnamefont {H.}~\bibnamefont
  {Nagaoka}},\ }\href@noop {} {\emph {\bibinfo {title} {Methods of information
  geometry}}},\ Vol.\ \bibinfo {volume} {191}\ (\bibinfo  {publisher} {American
  Mathematical Soc.},\ \bibinfo {year} {2000})\BibitemShut {NoStop}%
\end{thebibliography}%

\end{document}


\title{Supplemental Material for\\ Energy-information trade-off makes the cortical critical power law the optimal coding}

\author{Tsuyoshi Tatsukawa}
\author{Jun-nosuke Teramae}
\affiliation{%
 Graduate School of Informatics, Kyoto University, Sakyo-ku, Kyoto 606-8502, Japan\\
}

\maketitle

\section{Derivation of the probability density function\\ of neural activities for the power-law coding model}

In this section, we will explain how to derive the probability density function (Eq. (3) in the main text) of neural activities for the power-law coding model (Eq. (1) in the main text).

The probability density of neural activities can be obtained through the change of variables from the $2N+1$ Gaussian variables, $\eta$ and $\xi_i$ ${\left(i=1, \ldots, 2N\right)}$, to $2N$ neural variables, $r_i$ ${\left(i=1, \ldots, 2N\right)}$. To achieve this, we introduce an auxiliary variable $\phi = \theta + \eta$ representing the noisy input to the neurons and then marginalize it. For simplicity, we denote $r_{2n-1}=x_n$ and $r_{2n}=y_n$.
\begin{align}
 \begin{split}
  p(\bm{r}; \theta) &= p(\bm{x}, \bm{y}; \theta) \\
  &= \int_{-\infty}^{\infty} d\phi p(\bm{x}, \bm{y}, \phi; \theta) \\
  &= \int_{-\infty}^{\infty} d\phi p(\bm{\xi}, \eta; \theta)
  {\left\lvert\frac{\partial{\left(\bm{x}, \bm{y}, \phi\right)}}{\partial{\left(\bm{\xi}, \eta\right)}}\right\rvert}^{-1} \\
  &= \int_{-\infty}^{\infty} d\phi p_{\eta}(\eta; \theta) \prod_{i=1}^{2N} p_{\xi}(\xi_i)
  {\left\lvert\frac{\partial{\left(\bm{x}, \bm{y}, \phi\right)}}{\partial{\left(\bm{\xi}, \eta\right)}}\right\rvert}^{-1} \\
  &= \int_{-\infty}^{\infty} d\phi p_{\eta}(\phi - \theta) \prod_{n=1}^{N}
  p_{\xi}(x_n - n^{-\alpha / 2} \cos n \phi)
  p_{\xi}(y_n - n^{-\alpha / 2} \sin n \phi) \\
  &= \frac{1}{(2\pi\sigma_1^2)^{1/2}(2\pi\sigma_0^2)^N} \int_{-\infty}^{\infty} d\phi
  \exp {\left[ -\frac{1}{2\sigma_1^2}(\phi - \theta)^2
  - \frac{1}{2\sigma_0^2}\sum_{n=1}^N {\left({\left(x_n - n^{-\alpha/2}\cos{n\phi}\right)}^2
  + {\left(y_n - n^{-\alpha/2} \sin{n\phi}\right)}^2 \right)} \right]}.
 \end{split}
 \label{eq:p}
 \end{align}
Here, we have used the independence between Gaussian noise in the fourth line and the fact that the Jacobian satisfies ${\left\lvert \frac{\partial{\left(\bm{x}, \bm{y}, \phi\right)}}{\partial{\left(\bm{\xi}, \eta\right)}}\right\rvert} = 1$ in the fifth line. This gives the desired density function, Eq. (3) in the main text.

\section{Derivation of the Fisher information of the power-law coding for the case of one-dimensional input}

Here, we will derive the Fisher information (Eqs. (8) and (9) in the main text) of the power-law coding via the Gaussian approximation of the probability density function of the neural activities.

\subsection{Gaussian approximation of the probability density function}

Let us assume that the neural noise strength $\sigma_0$ and the input noise strength $\sigma_1$ are sufficiently small. Then, we can expand neural activity (Eq. (1) in the main text) and approximate its probability density \eqref{eq:p} as a multivariate Gaussian distribution, which corresponds to the second-order approximation of the exponent of the exponential function of \eqref{eq:p}.

The Gaussian distribution is characterized by the mean vector $\bm{m}$ and the covariance matrix $\Sigma$ of the neural activities $\bm{r}$. To derive their expressions, let us denote the neural activity as
\begin{align}
 \bm{r} = \bm{r}{\left(\theta+\eta\right)}+\bm{\xi},
\end{align}
where $r_{2n-1}{\left(\theta\right)}=n^{-\alpha/2}\cos{n\theta}$ and $r_{2n}{\left(\theta\right)}=n^{-\alpha/2}\sin{n\theta}$ in the current case. Then, the linear approximation gives
\begin{align}
 \bm{r}
 = \bm{r}{\left(\theta\right)}
 + \frac{\partial \bm{r}{\left(\theta\right)}}{\partial \theta} \eta
 + \bm{\xi}.
 \label{eq:rlin}
\end{align}
Therefore, by averaging this over the noise, we have the mean vector
\begin{align}
 \bm{m}
 = {\left\langle\bm{r}\right\rangle}_{\eta,\bm{\xi}}
 = \bm{r}{\left(\theta\right)}.
 \label{eq:m}
\end{align}
Now, let us introduce the susceptibility of the neural activity to input signal that is defined by the derivative of the mean $\bm{m}$ with respect to $\theta$,
\begin{align}
 \bm{\mu}
 = \frac{\partial \bm{m}}{\partial \theta}
 = \frac{\partial \bm{r}{\left(\theta\right)}}{\partial \theta}.
 \label{eq:mu}
\end{align}
Using the mean and the susceptibility, we can rewrite \eqref{eq:rlin} to
\begin{align}
 \bm{r}
 = \bm{m}
 + \bm{\mu} \eta
 + \bm{\xi}.
 \label{eq:rmmu}
\end{align}
This expression of the neural activity allows us to have the covariance matrix using the susceptibility as
\begin{align}
 \begin{split}
  \Sigma
  &=
  {\left\langle
  {\left( \bm{r} - \bm{m} \right)}
  {\left( \bm{r} - \bm{m} \right)}^\top
  \right\rangle}_{\eta, \bm{\xi}}
  \\
  &=
  {\left\langle
  {\left( \bm{\mu} \eta + \bm{\xi} \right)}
  {\left( \bm{\mu} \eta + \bm{\xi} \right)}^\top
  \right\rangle}_{\eta, \bm{\xi}}
  \\
  &=
  {\left\langle \bm{\xi} \bm{\xi}^\top \right\rangle}_{\bm{\xi}}
  +
  {\left\langle \eta^2 \right\rangle}_{\eta}
  \bm{\mu} \bm{\mu}^\top
  \\
  &=
  \sigma_0^2 \bm{I} + \sigma_1^2 \bm{\mu} \bm{\mu}^\top .
 \end{split}
 \label{eq:sigma}
\end{align}
Because \eqref{eq:sigma} give the relation between the fluctuation and the susceptibility of neural activity, we can regard this as a kind of {\bf the fluctuation-dissipation relation} for the neural coding.

This relation \eqref{eq:sigma} gives the interesting result that the susceptibility $\bm{\mu}$ is the eigenvector of the covariance matrix, and, in the current case, its eigenvalue is given by the generalized harmonic function $H_N(x) = \sum_{n=1}^N n^{-x}$ and, thus, by the Riemann zeta function $\zeta(x) = \sum_{n=1}^{\infty} n^{-x}$ in the limit of large numbers of neurons:
\begin{align}
 \Sigma\bm{\mu}
 = {\left(\sigma_0^2 \bm{I} + \sigma_1^2 \bm{\mu} \bm{\mu}^\top\right)} \bm{\mu}
 = {\left(\sigma_0^2 + \sigma_1^2 \bm{\mu}^\top \bm{\mu}\right)} \bm{\mu} = \lambda \bm{\mu},
 \label{eq:eigen}
\end{align}
where
\begin{align}
 \begin{split}
  \lambda
  &:= \sigma_0^2 + \sigma_1^2 \bm{\mu}^\top \bm{\mu} \\
  &= \sigma_0^2 + \sigma_1^2 \sum_{n=1}^N n^{2-\alpha}
  {\left(\sin^2(n \theta) + \cos^2(n \theta)\right)} \\
  &= \sigma_0^2 + \sigma_1^2 \sum_{n=1}^N n^{2-\alpha} \\
  &= \sigma_0^2 + \sigma_1^2 H_N(\alpha-2)
  \xrightarrow[N\to\infty]{} \sigma_0^2 + \sigma_1^2 \zeta(\alpha-2). \\
  \label{eq:lambda}
 \end{split}
\end{align}
Here, we have used
\begin{align}
 \mu_i
 = \frac{\partial r_i}{\partial \theta}
 =
 \begin{cases}
  - n^{1-\alpha / 2} \sin{(n \theta)} & (i=2n-1)\\
  n^{1-\alpha / 2} \cos{(n \theta)} & (i=2n)
 \end{cases}.
\end{align}
because $r_{2n-1}{\left(\theta\right)}=n^{-\alpha/2}\cos{n\theta}$ and $r_{2n}{\left(\theta\right)}=n^{-\alpha/2}\sin{n\theta}$ in our case. We will see soon that the equation \eqref{eq:eigen} and \eqref{eq:lambda} will play a key role in the derivation of the Fisher information.

By using the mean \eqref{eq:m} and the covariance \eqref{eq:sigma}, the Gaussian probability density function of the neural activity is given by
\begin{align}
 p(\bm{r}; \theta)
 = \frac{1}{\sqrt{(2\pi)^{2N}|\Sigma|}}
 \exp{{\left( - \frac{1}{2} (\bm{r}-\bm{m})^\top \Sigma^{-1}(\bm{r}-\bm{m}) \right)}},
 \label{eq:gauss_approx}
\end{align}
where $|\bm{A}|$ denotes the determinant of matrix $\bm{A}$ and
\begin{align}
 \Sigma^{-1}
 = \frac{1}{\sigma_0^2}{\left(I-\frac{\sigma_1^2}{\lambda}\bm{\mu} \bm{\mu}^\top\right)}
 \label{eq:inv}
\end{align}
is the inverse of the covariance matrix. Eq. \eqref{eq:inv} follows from a direct calculation using \eqref{eq:sigma} and \eqref{eq:eigen}.

\subsection{The fisher information of the power-law coding under the Gaussian approximation}

The density function \eqref{eq:gauss_approx} allows us to have the loglikelihood function of $\theta$ given the neural activities $\bm{r}$,
\begin{align}
 \begin{split}
  \log p(\bm{r}; \theta) &= - \frac{1}{2} (\bm{r}-\bm{m})^\top \Sigma^{-1} (\bm{r}-\bm{m}) + C \\
  &=  - \frac{1}{2 \sigma_0^2} (\bm{r}-\bm{m})^\top
  {\left(I-\frac{\sigma_1^2}{\lambda}\bm{\mu} \bm{\mu}^\top\right)}
  (\bm{r}-\bm{m}) + C \\
  &=  - \frac{1}{2 \sigma_0^2}
  {\left({\left|\bm{r}-\bm{m}\right|}^2
  -\frac{\sigma_1^2}{\lambda}{\left(\bm{\mu}^\top {\left(\bm{r}-\bm{m}\right)}\right)}^2\right)} + C,
  \label{eq:logp}
 \end{split}
\end{align}
where $C$ denotes the term not including the input stimulus $\theta$. Thus, the score function, i.e., the derivative of the loglikelihood with respect to the input $\theta$ is given by
\begin{align}
 \begin{split}
  \frac{\partial}{\partial\theta}\log p(\bm{r}; \theta)
  &= - \frac{1}{2\sigma_0^2} {\left(
  \frac{\partial}{\partial\theta}{\left|\bm{r}-\bm{m}\right|}^2
  - \frac{\sigma_1^2}{\lambda}
  \frac{\partial}{\partial\theta}
  {\left({\left(\bm{r}-\bm{m}\right)}^\top \bm{\mu}\right)}^2
  \right)} \\
  &= - \frac{1}{\sigma_0^2} {\left(
  - {\left(\bm{r}-\bm{m}\right)}^\top \frac{\partial\bm{m}}{\partial\theta}
  - \frac{\sigma_1^2}{\lambda}
  {\left(\bm{r}-\bm{m}\right)}^\top \bm{\mu}
  {\left(
  - \frac{\partial \bm{m}}{\partial\theta}^\top \bm{\mu}
  + {\left(\bm{r}-\bm{m}\right)}^\top \frac{\partial \bm{\mu}}{\partial\theta}
  \right)}
  \right)} \\
  &\approx
  \frac{1}{\sigma_0^2} {\left(
  {\left(\bm{r}-\bm{m}\right)}^\top \bm{\mu}
  - \frac{\sigma_1^2}{\lambda}
  {\left(\bm{r}-\bm{m}\right)}^\top \bm{\mu} \bm{\mu}^\top \bm{\mu}
  \right)} \\
  &=
  \frac{1}{\sigma_0^2} {\left(
  1 - \frac{\sigma_1^2}{\lambda}\bm{\mu}^\top \bm{\mu}
  \right)}
  {\left(\bm{r}-\bm{m}\right)}^\top \bm{\mu} \\
  &=
  \frac{1}{\lambda}
  {\left(\bm{r}-\bm{m}\right)}^\top \bm{\mu}. \\
 \end{split}
\end{align}
To obtain the third line, we used \eqref{eq:mu} and omitted the higher-order term of $\bm{r}-\bm{m}$ because it will just gives a higher-order correlation to the Fisher information that will vanish under the small noise approximation. The last line follows from the first line of \eqref{eq:lambda}.

By differentiating the score function again, we arrive at the Fisher information of the power-law coding
\begin{align}
 \begin{split}
  I{\left(\theta\right)}
  &=
  - {\left\langle
  \frac{\partial^2}{\partial \theta^2} \log p(\bm{r}; \theta)
  \right\rangle}_{\bm{r}} \\
  &=
  - \frac{1}{\lambda}
  {\left\langle
  \frac{\partial}{\partial \theta}{\left({\left(\bm{r}-\bm{m}\right)}^\top \bm{\mu}\right)}
  \right\rangle}_{\bm{r}} \\
  &=
  \frac{1}{\lambda}
  \bm{\mu}^\top
  \bm{\mu} \\
  &=
  \frac{\abs{\bm{\mu}}^2}{\sigma_0^2 + \sigma_1^2\abs{\bm{\mu}}^2} \\
  &=
  \frac{H_N{\left(\alpha-2\right)}}{\sigma_0^2 + \sigma_1^2 H_N{\left(\alpha-2\right)}}
  .
 \end{split}
 \label{eq:fi}
\end{align}
The last line follows from \eqref{eq:lambda}. In the limit of a large number of $N$, the fisher information converges to
\begin{align}
 \frac{\zeta{\left(\alpha-2\right)}}{\sigma_0^2 + \sigma_1^2 \zeta{\left(\alpha-2\right)}},
 \label{eq:fi_lim}
\end{align}
which is Eq. (8) in the main text. We used \eqref{eq:fi} to plot the dotted lines of Fig. 6 and used \eqref{eq:fi_lim} for the thick line.

\section{Derivation of the Fisher information of the power-law coding for high dimensional input stimulus}

In this section, extending the results of the above sections, we derive the Fisher information of the power-law coding for a multidimensional input stimulus.

To simplify the calculation, instead of using the trigonometric functions $\cos{n\theta}$ and $\sin{n\theta}$ ${\left(n=1,\cdots ,N\right)}$ to represent neural activity, let us introduce complex Fourier basis functions $e^{in\theta}/\sqrt{2}$ ${\left(n=\pm 1,\ldots,\pm N\right)}$ that satisfies the constraint $z_{-n}=\overline{z_n}$, where $\overline{z}$ is the complex conjugate of $z$. The factor $1/\sqrt{2}$ appears here to make the norm of the complex basis functions equal to that of the trigonometric functions in function space because it holds that
\begin{align*}
 \int_0^{2\pi}\left|\cos{n\theta}\right|^2 d\theta
 = \int_0^{2\pi}\left|\sin{n\theta}\right|^2 d\theta
 = \int_0^{2\pi}\left|\frac{1}{\sqrt{2}}e^{in\theta}\right|^2 d\theta .
\end{align*}
The neural activities for one dimensional input is thus rewritten by
\begin{align}
 r_n = z_n = \frac{1}{2^{1/2}}{\left|n\right|}^{-\alpha/2} e^{in{\left(\theta + \eta \right)}}+ \xi_n,
\end{align}
where $\xi_n$ is the complex Gaussian variable with the mean $0$ and the strength $\sigma_0$, namely, its real and imaginary parts independently follow the Gaussian distribution of the mean $0$ and the variance $\sigma_0^2/2$ with satisfying $\xi_{-n}=\overline{\xi_n}$, which gives
$
 {\left\langle\xi_{n} \xi_{-m}\right\rangle}
 = \delta_{nm}{\langle\xi_{n} \overline{\xi_{n}}\rangle}
 = \delta_{nm}{\langle{\left(\Re{\xi_{n}}\right)}^2+{\left(\Im{\xi_{n}}\right)}^2\rangle}
 = \sigma_0^2 \delta_{nm}.
$

Similarly, by regarding the neural representations of the $D$-dimensional input $\bm{\theta}={\left(\theta_1,\ldots \theta_D\right)}^\top$ as the multidimensional (complex) Fourier expansion, we can extend the above neural activity to
\begin{align}
 z_{\bm{k}}
 = \frac{1}{2^{D/2}} n{\left(\bm{k}\right)}^{-\alpha/2}
 e^{i \bm{k}^\top {\left(\bm{\theta} + \bm{\eta}\right)}}
 + \xi_{\bm{k}},
 \label{eq:z}
\end{align}
where each neuron is indexed by $D$-dimensional lattice vectors $\bm{k}={\left(k_1, \ldots ,k_D\right)}^\top$, where $k_d$ $(d=1,\ldots ,D)$ is the integer representing the frequency or the wavenumber of the neural activity for the $d$th input stimulus $\theta_d$. The function $n(\cdot)$ is a numbering that aligns the neurons in ascending order of their frequencies, and thus a function of the lattice vector $\bm{k}$. For now, we leave it as an arbitrary function satisfying the condition
\begin{align}
 n{\left(k_1,\ldots ,-k_d,\ldots ,k_D\right)}=n{\left(k_1,\ldots ,k_d,\ldots ,k_D\right)}
 \label{eq:n_symmetry}
\end{align}
for $d=1,\ldots ,D$, and will specify it later. Input noise $\bm{\eta}={\left(\eta_1,\ldots \eta_D\right)}^\top$ and neural noise $\xi_{\bm{k}}$ are a real and complex independent Gaussian variables, respectively, satisfying $\langle\eta_d\rangle=\langle\xi_{\bm{k}}\rangle=\langle\eta_d\xi_{\bm{k}}\rangle=0$, $\langle\eta_d \eta_k\rangle=\sigma_d^2\delta_{dk}$, and ${\langle\left(\Re{\xi_{\bm{k}}}\right)^2\rangle}={\langle\left(\Im{\xi_{\bm{k}}}\right)^2\rangle}=\sigma_0^2/2$, and thus,
\begin{align*}
 {\left\langle\xi_{\bm{k}} \xi_{-\bm{l}}\right\rangle}
 = \delta_{\bm{k}\bm{l}}{\left\langle\xi_{\bm{k}} \overline{\xi_{\bm{k}}}\right\rangle}
 = \delta_{\bm{k}\bm{l}}{\left\langle{\left(\Re{\xi_{\bm{k}}}\right)}^2+{\left(\Im{\xi_{\bm{k}}}\right)}^2\right\rangle}
 = \sigma_0^2 \delta_{\bm{k}\bm{l}},
\end{align*}
where $\delta_{\bm{k} \bm{l}}$ means $\delta_{k_1 l_1}\cdots\delta_{k_D l_D}$, and $\sigma_0$ and $\sigma_d$ are the strengths of the neural and $d$th input, respectively. The first equality of the above follows from $\xi_{\bm{-k}}=\overline{\xi_{\bm{k}}}$ that is required for $z_{-\bm{k}}=\overline{z_{\bm{k}}}$.

Similar to the previous sections, by assuming that the noise strengths are sufficiently, we approximate the density function of the neural activities with the multivariate (complex) Gaussian distribution
\begin{align}
 p{\left(\bm{z};\theta\right)}
 \propto \exp{\left(
 -\frac{1}{2} {\left(\bm{z}-\bm{m}\right)}^{*}
 \Sigma^{-1} {\left(\bm{z}-\bm{m}\right)}
 \right)}.
\end{align}
Here, we defined a column vector $\bm{z}$ by reshaping $z_{\bm{k}}=z_{k_1 \ldots k_D}$ that is indeed a tensor as indicated by the multiple subscripts. Thus, for instance,
\begin{align}
 \bm{z} =
 \begin{pmatrix}
  z_{-K_1,\ldots,-K_D} \\
  z_{-K_1,\ldots,-K_D+1} \\
  \vdots \\
  z_{K_1,\ldots,K_D}
 \end{pmatrix}.
\end{align}
The complex mean vector $\bm{m}$ and the complex covariance matrix $\Sigma$ of $\bm{z}$ are defined by
\begin{align}
 \bm{m} &= {\left\langle \bm{z} \right\rangle} \\
 \Sigma &= {\left\langle {\left(\bm{z}-\bm{m}\right)} {\left(\bm{z}-\bm{m}\right)}^{*} \right\rangle} ,
\end{align}
where $\bm{x}^{*}$ denotes the Hermitian conjugate of the vector $\bm{x}$. While $\bm{z}$ is a column vector as defined above, we continue to abuse notation of the tensor index $z_{\bm{k}}=z_{k_1 \ldots k_D}$ to denote its component as before for convenience. Similarly, we use $m_{\bm{k}}=m_{k_1 \ldots k_D}$ and $\Sigma_{\bm{k} \bm{l}}=\Sigma_{k_1 \ldots k_D, l_1 \ldots l_D}$ to denote components of the vector $\bm{m}$ and the matrix $\Sigma$, respectively.

The same argument as before gives the linear approximation of the complex neural activity as
\begin{align}
 \bm{z} = \bm{m} + \sum_{d=1}^{D} \bm{\mu}_d \eta_d + \bm{\xi},
\end{align}
where components of the complex mean $\bm{m}={\left(m_{\bm{k}}\right)}$ is given by
\begin{align}
 m_{\bm{k}}
 = {\left\langle z_{\bm{k}} \right\rangle}
 = \frac{1}{2^{D/2}}
 n{\left(\bm{k}\right)}^{-\alpha/2} e^{i \bm{k}^\top \bm{\theta}}
 \label{eq:m_multi}
\end{align}
and the susceptibility $\bm{\mu}_d$ $(d=1,\ldots ,D)$ is defined by the derivative the vector $\bm{m}$ with respect to the $d$th input signal $\theta_d$
\begin{align}
 \bm{\mu}_{d} := \frac{\partial \bm{m}}{\partial \theta_d}
 \label{eq:mu_m_multi}
\end{align}
whose components are given by
\begin{align}
 {\left(\bm{\mu}_{d}\right)}_{\bm{k}}
 = \frac{\partial m_{\bm{k}}}{\partial \theta_d}
 = i k_d m_{\bm{k}}
 = i k_d \frac{1}{2^{D/2}} n{\left(\bm{k}\right)}^{-\alpha/2}
  e^{i \bm{k}^\top \bm{\theta}} .
\end{align}
Then, using the susceptibility, we obtain an expression of the covariance as
\begin{align}
 \begin{split}
  \Sigma
  &=
  {\left\langle
  {\left( \bm{z} - \bm{m} \right)}
  {\left( \bm{z} - \bm{m} \right)}^{*}
  \right\rangle}_{\bm{\eta}, \bm{\xi}}
  \\
  &=
  {\left\langle
  {\left( \sum_{d=1}^{D} \bm{\mu}_d \eta_d + \bm{\xi} \right)}
  {\left( \sum_{d=1}^{D} \bm{\mu}_d \eta_d + \bm{\xi} \right)}^{*}
  \right\rangle}_{\bm{\eta}, \bm{\xi}}
  \\
  &=
  {\left\langle \bm{\xi} \bm{\xi}^{*} \right\rangle}_{\bm{\xi}}
  +
  \sum_{d=1}^{D}
  {\left\langle \eta_d^2 \right\rangle}_{\eta_d}
  \bm{\mu} \bm{\mu}^{*}
  \\
  &=
  \sigma_0^2 I + \sum_{d=1}^{D} \sigma_d^2 \bm{\mu}_d \bm{\mu}_d^{*} ,
 \end{split}
 \label{eq:sigma_multi}
\end{align}
which associates the fluctuation and the susceptibilities of the neural activity.

Because of the condition \eqref{eq:n_symmetry}, the susceptibilities satisfy the orthogonal condition
\begin{align}
 \begin{split}
  \bm{\mu}_{d}^{*} \bm{\mu}_{l}
  &=
  \frac{1}{2^D}
  \sum_{\bm{k} \in {\left\{ n{\left( \bm{k} \right)} \leq N \right\}} }
  k_d k_l n{\left(\bm{k}\right)}^{-\alpha}
  \\
  &=
  \delta_{dl}
  \frac{1}{2^D}
  \sum_{\bm{k} \in {\left\{ n{\left( \bm{k} \right)} \leq N \right\}} }
  k_d^2 n{\left(\bm{k}\right)}^{-\alpha}
  \\
  &=
  \delta_{dl} {\left|\bm{\mu}_{d}\right|}^2.
 \end{split}
 \label{eq:orthogonal}
\end{align}
Thus, we can show that the each susceptibility is an eigenvector of the complex covariance matrix
\begin{align}
 \Sigma \bm{\mu}_{d}
 =
 {\left(\sigma_0^2 I + \sum_{l=1}^{D} \sigma_l^2 \bm{\mu}_{l} \bm{\mu}_{l}^{*}\right)} \bm{\mu}_{d}
 =
 {\left(\sigma_0^2 + \sigma_d^2 {\left|\bm{\mu}_{d}\right|}^2\right)} \bm{\mu}_{d}
 =
 \lambda_d \bm{\mu}_{d},
 \label{eq:eigen_multi}
\end{align}
where we defined the $d$th eigenvalue by
\begin{align}
 \lambda_d
 &:=
 \sigma_0^2 + \sigma_d^2 {\left|\bm{\mu}_{d}\right|}^2 ,
 \label{eq:lambda_multi}
\end{align}
with
\begin{align}
 {\left|\bm{\mu}_{d}\right|}^2
 =
 \frac{1}{2^D}
 \sum_{\bm{k} \in {\left\{ n{\left( \bm{k} \right)} \leq N \right\}} }
 k_d^2 n{\left(\bm{k}\right)}^{-\alpha} .
 \label{eq:norm_mu}
\end{align}
Direct calculation using \eqref{eq:eigen_multi} gives the inverse of the covariance matrix as
\begin{align}
 \Sigma^{-1}
 = \frac{1}{\sigma_0^2}
 {\left(
 I - \sum_{d=1}^{D} \frac{\sigma_d^2}{\lambda_d} \bm{\mu}_d \bm{\mu}_d^{*}
 \right)},
 \label{eq:inv_multi}
\end{align}
and thus, the loglikelihood function of the multidimensional input $\bm{\theta}$ given the neural activities $\bm{z}$ is given by
\begin{align}
 \begin{split}
  \log p(\bm{z}; \bm{\theta})
  &=
  - \frac{1}{2} (\bm{z}-\bm{m})^{*} \Sigma^{-1} (\bm{z}-\bm{m}) + C
  \\
  &=
  - \frac{1}{2\sigma_0^2} (\bm{z}-\bm{m})^{*}
  {\left(
  I - \sum_{d=1}^{D} \frac{\sigma_d^2}{\lambda_d} \bm{\mu}_{d} \bm{\mu}_d^{*}
  \right)}
  (\bm{z}-\bm{m}) + C
  \\
  &=
  - \frac{1}{2\sigma_0^2}
  {\left(
  {\left|\bm{z}-\bm{m}\right|}^2
  - \sum_{d=1}^{D} \frac{\sigma_d^2}{\lambda_d}
  {\left|{\left(\bm{z}-\bm{m}\right)}^{*}\bm{\mu}_{d}\right|}^2
  \right)} + C .
 \end{split}
\end{align}
where $C$ denotes the term not including $\bm{\theta}$. The derivative of the loglikelihood with respect to a component of the input vector gives the score function
\begin{align}
 \begin{split}
  \frac{\partial}{\partial \theta_i} \log p(\bm{z}; \bm{\theta})
  &=
  - \frac{1}{2\sigma_0^2}
  {\left(
  -\frac{\partial \bm{m}^{*}}{\partial \theta_i}
  {\left(\bm{z}-\bm{m}\right)}
  - \sum_{d=1}^{D} \frac{\sigma_d^2}{\lambda_d}
  {\left(
  -\frac{\partial \bm{m}^{*}}{\partial \theta_i} \bm{\mu}_d
  + {\left(\bm{z}-\bm{m}\right)}^{*} \frac{\partial \bm{\mu}_d}{\partial \theta_i}
  \right)}
  \bm{\mu}_d^{*} {\left(\bm{z}-\bm{m}\right)}
  + \rm{h.c.}
  \right)} \\
  &\approx
  \frac{1}{2\sigma_0^2}
  {\left(
  \bm{\mu}_i^{*}
  {\left(\bm{z}-\bm{m}\right)}
  - \sum_{d=1}^{D} \frac{\sigma_d^2}{\lambda_d}
  \bm{\mu}_i^{*} \bm{\mu}_d
  \bm{\mu}_d^{*} {\left(\bm{z}-\bm{m}\right)}
  + \rm{h.c.}
  \right)} \\
  &=
  \frac{1}{2\sigma_0^2}
  {\left(
  1 - \frac{\sigma_i^2 {\left|\bm{\mu}_i\right|}^2}{\lambda_i}
  \right)}
  \bm{\mu}_i^{*}
  {\left(\bm{z}-\bm{m}\right)}
  + \rm{h.c.} \\
  &=
  \frac{1}{2\lambda_i}
  \bm{\mu}_i^{*}
  {\left(\bm{z}-\bm{m}\right)}
  + \rm{h.c.},
 \end{split}
\end{align}
where h.c. denotes the Hermitian conjugate of previous terms. To obtain the second line, we used \eqref{eq:mu_m_multi} and omitted the second-order term of $\bm{z}-\bm{m}$ that will only give higher-order terms to the Fisher information which vanish anyway under the small noise assumption. The third line is from the orthogonality \eqref{eq:orthogonal} and the last line follows from the definition of the eigenvalue \eqref{eq:lambda_multi}.

Then, we obtain the component of the Fisher information by averaging the negative derivative of the score function
\begin{align}
 \begin{split}
  I_{ij}{\left(\bm{\theta}\right)}
  &=
  {\left\langle
  -\frac{\partial^2}{\partial \theta_i \partial \theta_j } \log p(\bm{z}; \bm{\theta})
  \right\rangle}_{\bm{z}}
  \\
  &=
  {\left\langle
  -\frac{\partial}{\partial \theta_j }
  {\left(
  \frac{1}{2\lambda_i}
  \bm{\mu}_i^{*}
  {\left(\bm{z}-\bm{m}\right)}
  + \rm{h.c.}
  \right)}
  \right\rangle}_{\bm{z}}
  \\
  &=
  \frac{1}{\lambda_i} \bm{\mu}_i^{*} \bm{\mu}_j
  \\
  &=
  \frac{{\left|\bm{\mu}_i\right|}^2}{\lambda_i} \delta_{ij}
  \\
  &=
  \frac{{\left|\bm{\mu}_i\right|}^2}{\sigma_0^2+\sigma_i^2 {\left|\bm{\mu}_i\right|}^2} \delta_{ij}.
 \end{split}
 \label{eq:fi_multi}
\end{align}
Note that \eqref{eq:fi_multi} with \eqref{eq:norm_mu} recovers the Fisher information \eqref{eq:fi} for the case one-dimensional input as a special case,
\begin{align}
 I_{11}{\left(\theta\right)}
 =
 \frac{{\left|\bm{\mu}_1\right|}^2}{\sigma_0^2+\sigma_1^2 {\left|\bm{\mu}_1\right|}^2}
 =
 \frac{\frac{1}{2}\sum_{k=-N}^N k^2 k^{-\alpha}}
 {\sigma_0^2+\sigma_1^2 \frac{1}{2}\sum_{k=-N}^N k^2 k^{-\alpha}}
 =
 \frac{\sum_{k=1}^N k^2 k^{-\alpha}}
 {\sigma_0^2+\sigma_1^2 \sum_{k=1}^N k^2 k^{-\alpha}}
 =
 \frac{H_N{\left(\alpha-2\right)}}
 {\sigma_0^2 + \sigma_1^2 H_N{\left(\alpha-2\right)}}.
\end{align}

\begin{figure}
 \centering
 \includegraphics{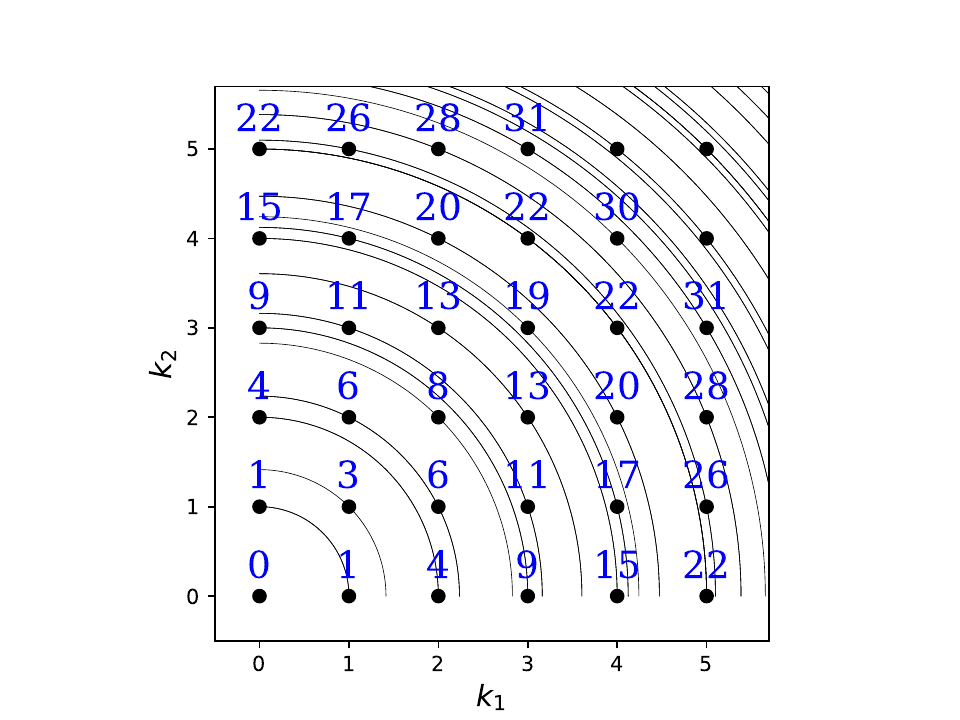}
 \caption{An example of the numbering $n(\bm{k})$ that aligns the neurons in ascending order of their frequency $\bm{k}$ for the case of $D=2$. The numbering function, or the rank precisely, $n(\bm{k})$ of the neuron at the lattice $\bm{k}$ is given as the numbers of neurons, or the lattices, whose distance from the origin is smaller than that of the lattice $\bm{k}$ (blue numbers). Therefore, the function $n(\bm{k})$ is given by the number of lattices inside the circle with radius $r(\bm{k})$ (black lines), whose leading order term is equal to the area of the disk enclosed by the circle, i.e., 2-ball, for large $n(\bm{k})$.}
 \label{fig:figS1}
\end{figure}

To proceed further, we need to specify the function $n{\left(\bm{k}\right)}$ that ranks the multidimensional lattice vectors $\bm{k}={\left(k_1,\ldots , k_D\right)}$ in ascending order of $\bm{k}$. Since ``ascending order'' is ambiguous in the multidimensional lattice space, we define the rank as it is consistent with the distance from the origin $r{\left(\bm{k}\right)}=\sqrt{k_1^2+\cdots +k_D^2}$. For example, we can define the function by
\begin{align}
 n{\left(\bm{k}\right)}
 =
 \sum_{\bm{h}\geq 0}
 \mathbbm{1}_{r{\left(\bm{h}\right)} < r{\left(\bm{k}\right)}} {\left[\bm{h}\right]}
 =
 \frac{1}{2^D}
 \sum_{\bm{h}}
 \mathbbm{1}_{r{\left(\bm{h}\right)} < r{\left(\bm{k}\right)}} {\left[\bm{h}\right]},
 \label{eq:n}
\end{align}
where $\bm{h}\geq 0$ means $h_d\geq 0$ for all $d$, and $\mathbbm{1}_A{\left[\bm{h}\right]}$ is the indicator function that returns one for the lattice point $\bm{h}$ satisfying the condition $A$ and zero otherwise. Thus, the definition above means that $n$ of the lattice point $\bm{k}$ is defined as the number of lattices $\bm{h}(\geq 0)$ whose distance from the origin is smaller than that of $\bm{k}$. In other words, $n(\bm{k})$ is equal to the number of lattice points inside the open ball with radius $r{\left(\bm{k}\right)}$ and in the non-negative orthant. Figure \ref{fig:figS1} illustrates this for the case of $D=2$. Note that the value of $n(\bm{k})$ for $\bm{k}\ngeq 0$ is determined by the symmetry condition \eqref{eq:n_symmetry}.

Because the sum of the right hand side of \eqref{eq:n} is approximated by the volume of the D-ball of radius $r$ for large $n$, we have
\begin{align}
 n{\left(\bm{k}\right)} =  \frac{1}{2^D} V_D r{\left(\bm{k}\right)}^D,
\end{align}
where $V_D = \pi^{D/2}/\Gamma {\left(D/2+1\right)}$ is the volume of the unit D-ball. It gives
\begin{align}
 k_1^2+\cdots + k_D^2
 = r{\left(\bm{k}\right)}^2
 = 4 {\left(\frac{1}{V_D}n{\left(\bm{k}\right)}\right)}^{2/D}.
\end{align}
Then using the symmetry of the function $n{\left(\bm{k}\right)}$, we have
\begin{align}
 \begin{split}
  {\left|\bm{\mu}_1\right|^2} = \cdots = {\left|\bm{\mu}_D\right|^2}
  &=
  \frac{1}{D}
  {\left({\left|\bm{\mu}_1\right|^2} + \cdots + {\left|\bm{\mu}_D\right|^2}
  \right)}
  \\
  &=
  \frac{1}{D 2^D}
  \sum_{\bm{k} \in {\left\{ n{\left( \bm{k} \right)} \leq N \right\}} }
  {\left(k_1^2+\cdots + k_D^2\right)} n{\left(\bm{k}\right)}^{-\alpha}
  \\
  &=
  \frac{1}{D 2^D}
  \sum_{\bm{k} \in {\left\{ n{\left( \bm{k} \right)} \leq N \right\}} }
  r{\left(\bm{k}\right)}^2 n{\left(\bm{k}\right)}^{-\alpha}
  \\
  &=
  \frac{1}{D}
  \sum_{\bm{k} \in {\left\{\bm{k} \geq 0 \land n{\left( \bm{k} \right)} \leq N \right\}} }
  r{\left(\bm{k}\right)}^2 n{\left(\bm{k}\right)}^{-\alpha}
  \\
  &=
  \frac{4}{D} V_D^{-2/D}
  \sum_{\bm{k} \in {\left\{\bm{k} \geq 0 \land n{\left( \bm{k} \right)} \leq N \right\}} }
  n{\left(\bm{k}\right)}^{-\alpha+2/D}
  \\
  &=
  \frac{4}{D} V_D^{-2/D}
  \sum_{n=1}^N n^{-\alpha+2/D}
  \\
  &=
  \frac{4}{D} V_D^{-2/D}
  H_N{\left(\alpha-2/D\right)}.
 \end{split}
\end{align}
By putting this to \eqref{eq:fi_multi}, we arrive at
\begin{align}
 I_{ij}{\left(\bm{\theta}\right)}
 =
 \frac{H_N{\left(\alpha-2/D\right)}}
 {\sigma_0^2 D V_D^{2/D} / 4 + \sigma_i^2 H_N{\left(\alpha-2/D\right)}} \delta_{ij},
 \label{eq:fi_multi_j}
\end{align}
which converges to
\begin{align}
 \frac
 {\zeta{\left(\alpha-2/D\right)}}
 {\sigma_0^2 D V_D^{2/D} / 4 + \sigma_i^2 \zeta{\left(\alpha-2/D\right)}},
 \label{eq:fi_multi_n_inf}
\end{align}
in the limit of $N\to\infty$, which further converges to
\begin{align}
 \frac
 {\zeta{\left(\alpha\right)}}
 {e \pi \sigma_0^2 / 2+ \sigma_i^2 \zeta{\left(\alpha\right)}},
 \label{eq:fi_multi_nd_inf}
\end{align}
in the limit of $D\to\infty$ because it holds that
\begin{align}
 \begin{split}
  \frac{D V_D^{2/D}}{4}
  =
  \frac{D}{4}{\left(\frac{\pi^{D/2}}{\Gamma{\left(D/2+1\right)}}\right)}^{2/D}
  \approx
  \frac{D}{4}{\left(\frac{\pi^{D/2}}{{\left(\pi D\right)}^{1/2}{\left(\frac{D}{2e}\right)}^{D/2}}\right)}^{2/D}
  =
  \frac{e \pi}{2{\left(\pi D\right)}^{1/D}}
  \xrightarrow[D\to\infty]{}
  \frac{e \pi}{2}
 \end{split}
\end{align}
due to the Stirling's formula. These expressions of the Fisher information \eqref{eq:fi_multi_n_inf} and \eqref{eq:fi_multi_nd_inf} are Eqs. (9) and (10) in the main text, respectively.

\section{Condition for the regularization parameter of the energy-aware performance measure}

For completeness, in this section we provide the condition that the energy cost given by the second term of the energy-aware performance measure $J_D{\left(\alpha\right)}$ (Eq. (11) in the main text) does not overwhelm the first term.

The energy-aware performance measure was given by
\begin{align}
 \begin{split}
  J_D{\left(\alpha\right)}
  =
  I_D{\left(\alpha\right)} - \gamma \zeta{\left(\alpha\right)}
  =
  \frac
  {\zeta(\alpha-2/D)}
  {\sigma_0^2 D V_D^{2/D} / 4 + \sigma_i^2 \zeta(\alpha-2/D)}
  - \gamma \zeta{\left(\alpha\right)}.
 \end{split}
\end{align}
Considering that the second term $-\gamma \zeta\left(\alpha\right)$ is a monotonically increasing function of $\alpha$ whereas the first term of the Fisher information is constant until it monotonically decreases from the critical point $\alpha=\alpha_c$, it is obvious that the $J_D{\left(\alpha\right)}$ is monotonically increasing for $\alpha<\alpha_c=1+2/D$. Therefore, the condition that $\alpha_c$ gives the maximum of $J_D$ is given by the one that the derivative of $J_D{\left(\alpha\right)}$ at $\alpha\to\alpha_c+0$ is negative:
\begin{align}
 \lim_{\alpha \to \alpha_c + 0} J'_D{\left(\alpha\right)}
 &=
 I'_D{\left(\alpha_c + 0\right)} - \gamma \zeta'(\alpha_c)
 \\
 &=
 \frac{\sigma_0^2 D V_D^{2/D} / 4 \zeta'(1)}
 {{\left(\sigma_0^2 D V_D^{2/D} / 4 + \sigma_i^2 \zeta(1) \right)}^2}
 -
 \gamma \zeta'(\alpha_c)
 < 0 .
\end{align}
By Solving this we obtain the condition for the regularization parameter $\gamma$ as
\begin{align}
 \begin{split}
  \gamma
  &<
  - \frac{\sigma_0^2 D V_D^{2/D} / 4 \zeta'(1)}
  {{\left(\sigma_0^2 D V_D^{2/D} / 4 + \sigma_i^2 \zeta(1) \right)}^2 \zeta'(\alpha_c)}
  \\
  &=
  - \frac{\sigma_0^2 D V_D^{2/D}}{4\sigma_i^4\zeta'(\alpha_c)} =: \gamma_c,
 \end{split}
\end{align}
where we used the formula $\zeta'(1)/{\left(a + \zeta(1)\right)}^2=-1$ that follows from the fact that the Riemann zeta function $\zeta(z)$ has only a simple pole at $z=1$ with residual $1$, i.e., the principal part of $\zeta(z)$ is $1/(1-z)$.
Conversely, when $\gamma$ satisfies the above condition, we can easily check that $J_D'{\left(\alpha\right)}<0$ for all $\alpha>\alpha_c$ by using the fact that $I_D'{\left(\alpha\right)}$ is an increasing function there,
\begin{align}
 \begin{split}
  J_D'{\left(\alpha\right)}
  &=
  I_D'{\left(\alpha\right)}
  -
  \gamma \zeta'(\alpha)
  <
  I_D'{\left(\alpha\right)}
  -
  I_D'{\left(\alpha_c\right)}
  <
  0.
 \end{split}
\end{align}
We used $\gamma=\gamma_c / 10$ for Fig. 4 in the main text.
